\documentclass[10pt,final,journal]{IEEEtran}
\usepackage{cite}
\usepackage{url}
\usepackage[dvipsnames]{xcolor}
\usepackage{bbding}
\usepackage{hyperref}

\usepackage{multirow} 
\usepackage{amsmath}
\usepackage{booktabs}
\usepackage{makecell}
\usepackage{amsmath}
\usepackage{amssymb}
\usepackage{times}
\usepackage{epsfig}
\usepackage{graphicx}
\usepackage{float}

\usepackage{pifont}
\usepackage{color}
\usepackage{cite}
\usepackage{xcolor,graphicx,float}
\graphicspath{{./figures/}}
\usepackage[numbers,sort&compress]{natbib}
\usepackage{amsmath}
\usepackage{bm}
\usepackage{soul}

\newcommand{\ie}{\textit{i}.\textit{e}.}
\newcommand{\eg}{\textit{e}.\textit{g}.}
\usepackage{ragged2e}

\ifCLASSOPTIONcompsoc
\usepackage[caption=false,font=normalsize,labelfont=sf,textfont=sf]{subfig}
\else
\usepackage[caption=false,font=footnotesize]{subfig}
\fi

\begin{document}

%

\title{Improving the Stability and Efficiency of Diffusion Models for Content Consistent Super-Resolution}
%
%

\author{Lingchen~Sun, Rongyuan Wu,  Jie~Liang,
        Zhengqiang Zhang, Hongwei Yong, Lei~Zhang,~\IEEEmembership{Fellow,
        ~IEEE} 
\thanks{L.~Sun, R.~Wu, Z.~Zhang, H. Yong ,and  L.~Zhang are with the Department of Computing, the Hong Kong Polytechnic University, Hong Kong (e-mail: ling-chen.sun@connect.polyu.hk;
rong-yuan.wu@connect.polyu.hk;
zhengqiang.zhang@connect.polyu.hk;
hongwei.yong@polyu.edu.hk;
cslzhang@comp.polyu.edu.hk). J.~Liang is with the OPPO Research Institute (e-mail: liang27jie@gmail.com). This work is supported by the Hong Kong RGC RIF grant (R5001-18) and the PolyU-OPPO Joint Innovation Lab.}}
\maketitle

\begin{abstract}
The generative priors of pre-trained latent diffusion models (DMs) have demonstrated great potential to enhance the visual quality of image super-resolution (SR) results. 
However, the noise sampling process in DMs introduces randomness in the SR outputs, and the generated contents can differ a lot with different noise samples. The multi-step diffusion process can be accelerated by distilling methods, but the generative capacity is difficult to control.
To address these issues, we analyze the respective advantages of DMs and generative adversarial networks (GANs) and propose to partition the generative SR process into two stages, where the DM is employed for reconstructing image structures and the GAN is employed for improving fine-grained details. 
Specifically, we propose a non-uniform timestep sampling strategy in the first stage. A single timestep sampling is first applied to extract the coarse information from the input image, then a few reverse steps are used to reconstruct the main structures. 
In the second stage, we finetune the decoder of the pre-trained variational auto-encoder by adversarial GAN training for deterministic detail enhancement. Once trained, our proposed method, namely content consistent super-resolution (CCSR),
allows flexible use of different diffusion steps in the inference stage without re-training. 
Extensive experiments show that with 2 or even 1 diffusion step, CCSR can significantly improve the content consistency of SR outputs while keeping high perceptual quality. Codes and models can be found at \href{https://github.com/csslc/CCSR}{https://github.com/csslc/CCSR}.
\end{abstract}


%
\IEEEpeerreviewmaketitle

\section{Introduction}
\label{sec:intro}
%
%
%
%

\begin{figure*}
	\centering 
	\includegraphics[scale=1.0]{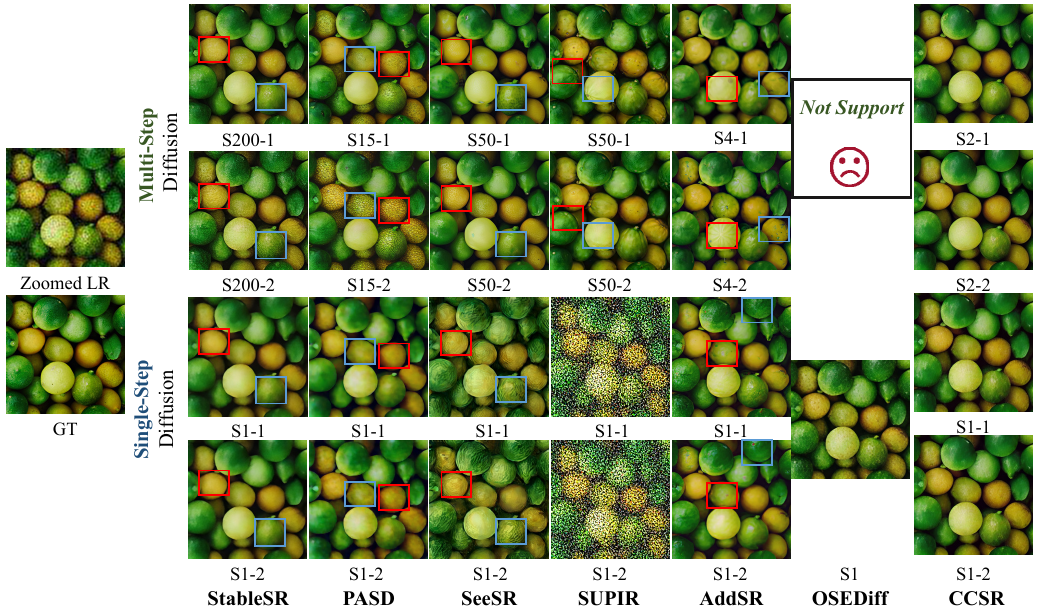}

	\caption{Visual comparisons between the super-resolution outputs with the same input low-quality image but two different noise samples by different DM-based methods. $S$ denotes diffusion sampling timesteps. Please zoom in for a better view. Existing DM-based methods, including StableSR \cite{stablesr}, PASD \cite{pasd}, SeeSR \cite{seesr}, SUPIR \cite{supir} and AddSR \cite{AddSR}, show noticeable instability with the different noise samples. OSEDiff \cite{osediff} directly takes low-quality image as input withour noise sampling. It is deterministic and stable, but cannot perform multi-step diffusion for high generative capacity. In contrast, our proposed CCSR method is flexible for both multi-step diffusion and single-step diffusion, while producing stable results with high fidelity and visual quality.}
	\label{fig1}

\end{figure*}

\IEEEPARstart{I}mage super-resolution (SR) aims to recover a high-resolution (HR) image with better visual quality from its low-resolution (LR) observation, which is a typical ill-posed problem \cite{9044873}. Many of the previous deep learning-based SR methods \cite{srgan,esrgan}, including those convolutional neural networks (CNN) \cite{srcnn, haris2018deep, kim2016deeply} and Transformer \cite{swinir, elan, hat2023} based ones, focus on the network backbone design by assuming simple and known image degradation, \eg, bicubic down-sampling, down-sampling after Gaussian blur. Though great progress has been achieved in simulated inputs, they may fail to restore realistic and rich details when facing images with complex and unknown degradations in real-world applications. 

To solve the problem, some methods have been proposed to improve training pairs to better fit the degradation in real applications, for example, by collecting real-world LR-HR pairs \cite{realsr, drealsr} or simulating more complex and comprehensive degradations \cite{realesrgan, bsrgan}. Besides the training data, training losses and strategies also play key roles in generating realistic details. Pixel-wise losses like $\ell_1$ and MSE are prone to generating over-smoothed details \cite{LDL}. The SSIM loss \cite{ssim} and perceptual loss \cite{perceptualloss} can alleviate this issue to some extent, while the adversarial loss from the generative adversarial network (GAN) provides a  more effective solution to reproduce richer and more realistic SR details \cite{srgan,esrgan, realesrgan, LDL, DASR}. Specifically, GAN-based methods perform favorably in reconstructing some specific scenarios such as face \cite{face1} due to the relatively small space. However, when handling natural images, GAN can hardly ensure good guesses on image structures due to its limited prior modeling capacity on natural scenes \cite{LDL,zhang2022perception, chen2023human, eaadam}, resulting in unpleasant visual artifacts.

Recently, the Denoising Diffusion Probabilistic Model (DDPM) \cite{ddpm} and its variants \cite{ddim, dpm} have achieved unprecedented successes in numerous fields \cite{diffusionshadow, sr3}. Compared to GANs, diffusion models (DMs) can learn richer natural image priors, which can be used for improving image restoration performance \cite{diffusion-restoration}. 
By using the LR image as a condition, some recent works \cite{stablesr,diffbir,pasd,seesr,supir, mei2023conditional, AddSR} have exploited the natural image priors in pre-trained text-to-image DMs \cite{stablediffusion} for more realistic SR.
In general, some methods \cite{stablesr,diffbir,pasd,seesr,supir} leverage a number of noise sampling steps to reconstruct image semantic structures and fine details. However, the noise sampling process also introduces randomness in the SR outputs so that the generated contents with different noise samples can vary a lot. The top rows in Fig. \ref{fig1} show an example. With the same LR as input, we run StableSR \cite{stablesr}, PASD \cite{pasd}, SeeSR \cite{seesr} and SUPIR \cite{supir} two times with two different noise seeds. We can see that while DM-based SR methods can generate rich details, their outputs in different runs may differ from each other, especially in the textures and details. Additionally, DM-based methods may produce unfaithful and visually over-enhanced or blurry details compared to the input and the ground truth. Such kind of instability significantly affects the fidelity and content consistency of SR outputs. 
 
 Directly reducing the number of sampling steps can mitigate the instability of DM-based SR results, but it also leads to deteriorated visual generation performance. As shown in the bottom rows of Fig. \ref{fig1}, single-step StableSR outputs blurry images, single-step SeeSR generates visually unappealing details, and single-step SUPIR is unable to remove noise and generate the desired image. AddSR \cite{AddSR} leverages distillation approaches to accelerate the diffusion process for SR tasks. It can maintain strong generative capability with fewer steps, but it tends to produce unstable and unfaithful results with a single step. The recently developed one-step diffusion method OSEDiff \cite{osediff} directly takes the LR as input without noise sampling. Its output is deterministic and stable. However, its generation capacity is limited since it cannot perform multi-step diffusion. 
 It is highly demanded to develop a flexible DM-based SR method that can achieve stable and visually pleasing results with both multi-step and single-step diffusion reverse sampling, meeting different perception-fidelity balanced requirements.

\begin{figure*}[ht]
	\centering 

	\includegraphics[scale=0.6]{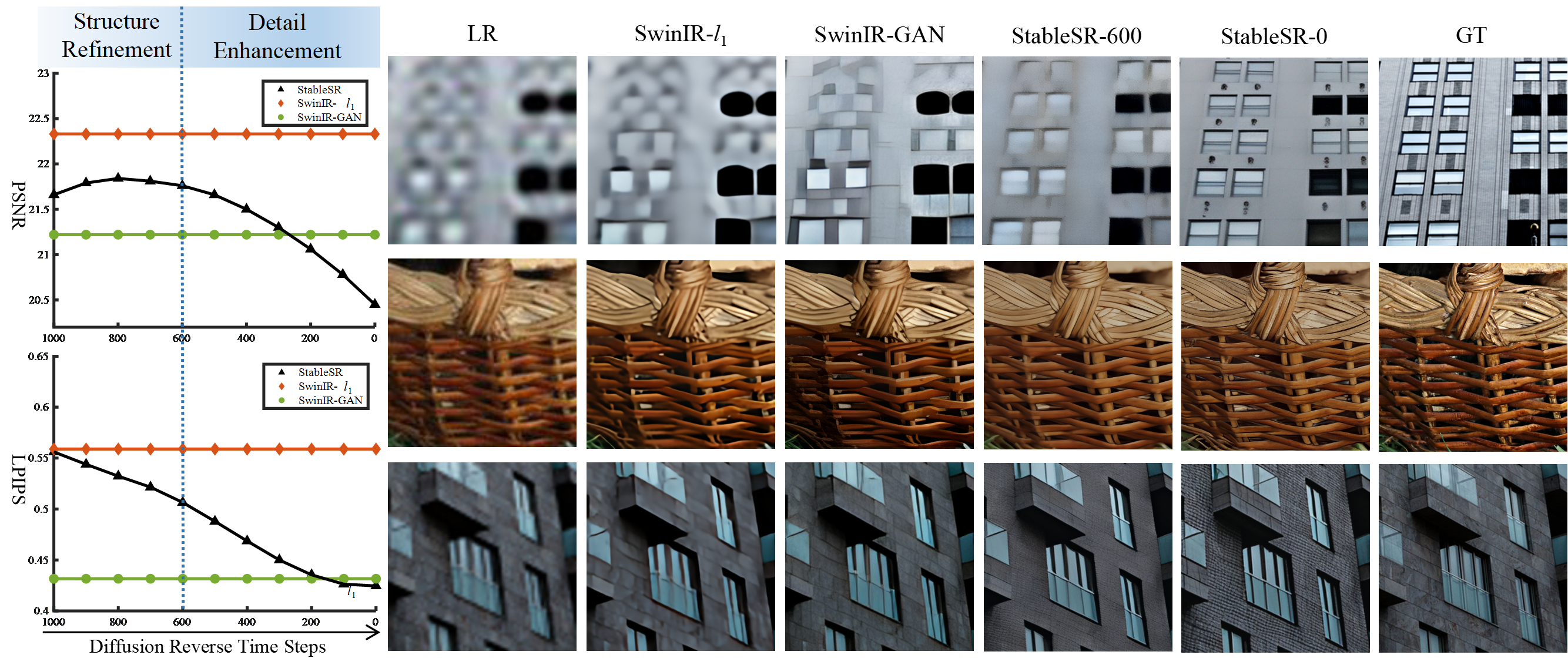}
	\caption{\textbf{Left}: PSNR and LPIPS indices of SR outputs by SwinIR-$\ell_1$, SwinIR-GAN \cite{swinir} and StableSR \cite{stablesr} at different steps on the DIV2K dataset. 
 \textbf{Right}: Visual comparisons of the SR results on three LR images of different quality levels. 
 Please refer to Section \ref{sec:motivation} for detailed explanations of this figure.}

	\label{fig2}
\end{figure*}

To achieve the goal mentioned above, we propose a Content Consistent Super-Resolution (CCSR) approach in this paper, which leverages diffusion priors to reproduce image structures that are faithful to the LR input, and employs GAN for subsequent detail and texture enhancement. Our method is inspired by the observation that DM is powerful in generating object structures. At the same time, GAN can effectively synthesize fine-grained details once the main structures are reconstructed, as shown in Fig. \ref{fig2}. Therefore, we partition the SR process into two stages to maximize the advantages of DM and GAN  in structure generation and detail synthesis. In the first stage, we propose a non-uniform timestep sampling strategy by using DM. A few timesteps are employed to generate a clearer image structure, after which the intermediate diffusion steps are truncated to avoid generating unfaithful details \cite{rissanen2022generative, choi2022perception}.
If efficiency is of particular concern, we can design a single reverse timestep, instead of progressive generation, for structure extraction since most of the low-frequency information can be obtained from the LR input. In the second stage, we finetune the pre-trained VAE decoder \cite{vae} with adversarial GAN training. The input to this stage is the output from the first stage. Therefore, the finetuned VAE decoder can accomplish both latent feature decoding and detail enhancement simultaneously without introducing additional computation burden. Once trained, during inference, our CCSR model allows the use of either single step or multi-step diffusion for HR image synthesis. This flexibility enable us to achieve diverse perception-distortion balances based on different user preferences. 

To sum up, we first analyze the instability of DM-based SR methods. Then, we propose CCSR, which disentangles the SR process into structure generation and detail refinement. Extensive experiments show that the proposed CCSR can improve both the content consistency and visual quality of the SR outputs, as shown in Fig. \ref{fig1}. In addition, CCSR supports multi-timestep and single-timestep sampling simultaneously, which is more flexible than previous methods to balance efficiency and generation capacity. 



\section{Related Work}
\subsection{Image Super-Resolution}
Traditional deep learning-based SR methods are designed for better image fidelity measures such as PSNR and SSIM \cite{ssim} indices. The earliest representative works include SRCNN \cite{srcnn} and DnCNN \cite{dncnn}. After that, various novel strategies, such as dense \cite{denseSR}, residual \cite{zhang2018residual} and recursive connections \cite{kim2016deeply} and non-local networks \cite{wang2021lightweight}, attention mechanism\cite {swinir, elan, hat2023} have been proposed to improve the SR performance. 
To improve the quality of real-world LR images with complex and even unknown degradations, researchers have collected the real-world LR-HR paired datasets \cite{realsr,drealsr} to train the network, or simulated real-world degradations using elegantly designed procedures \cite{bsrgan, realesrgan}. BSRGAN \cite{bsrgan} simulates real-world degradations by using a random shuffling strategy of basic degradation operators, while RealESRGAN \cite{realesrgan} uses a high-order degradation modeling by repeatedly applying a series of degradation operations. Subsequently, many SR methods apply GANs with the elaborated loss functions \cite{DASR, LDL, chen2023human} to handle real-world degradations. In general, for stable training, the $\ell_1$ loss is firstly applied to extract coarse structure information from LR, and then the GAN is used for enhancing details. GAN-based models yield sharper lines and more high-frequency details. However, the performance highly relies on the structure restored by $\ell_1$ loss. Based on the inaccurate structure, GAN struggles to reproduce rich and natural details.
Due to their powerful image priors, the recently developed DMs provide an alternative to GANs for solving the SR task.

\subsection{Diffusion SR Models}
Recently, the generative DMs \cite{ddpm, ddim, stablediffusion} have been rapidly developed, which can learn richer natural image priors than GAN, and DM priors have been successfully employed for image SR tasks \cite{ddnm, ddrm,ldmsr, resshift,stablesr, diffbir, pasd, seesr, supir}. 
There are three main types of DM-based SR methods. The first type \cite{ddnm, ddrm, gdp} modifies the reverse transition of a pre-trained DM using gradient descent. These methods are training-free but assume a pre-defined image degradation model. The second type \cite{ldmsr, resshift} retrains a DM from scratch on the paired training data. 
The third type \cite{stablesr, diffbir, pasd, mei2023conditional,seesr, supir, AddSR, osediff} leverages the strong image priors of large-scale pre-trained DM, such as the text-to-image models \cite{stablediffusion}, and introduces an adapter \cite{controlnet, t2i, lora} to fine-tune them. With the LR image as the control signal, high-quality SR outputs can be obtained. Some methods \cite{pasd, seesr, AddSR, osediff} introduce additional high-level models \cite{YOLO,RAM,liu2024visual,zhang2024dual,li2024univs,li2023opensd} to incorporate semantic information into the DM process. However, these multi-step methods suffer from the inconsistency and instability of SR results due to the randomness of DMs. In addition, some methods \cite{AddSR, osediff} distill few-step efficient models from multi-step models, but they are difficult to control the generative capacity.

\section{The Proposed Method}

\subsection{Motivation and Framework}
\label{sec:motivation}

Let's first investigate how the structures and details are generated by GAN and DM-based SR methods at different stages. 
In the left part of Fig. \ref{fig2}, we plot the PSNR and LPIPS indices of SR outputs by SwinIR-$\ell_1$, SwinIR-GAN \cite{swinir} and StableSR \cite{stablesr} at different timesteps on the DIV2K dataset.
For the GAN-based SwinIR-GAN \cite{swinir}, the $\ell_1$ loss is firstly applied to extract the information from LR input to ensure fidelity, and then the adversarial GAN loss is used for enhancing details \cite{srgan}. Therefore, the fidelity-based PSNR (the larger the better) and perception-based LPIPS (the lower the better) indices show rather different trends for SwinIR-GAN. For the DM-based method StableSR \cite{stablesr}, the image structures are reconstructed in the early diffusion stages, leading to an increase in PSNR. In the later diffusion stages, the gradually synthesized details lead to a significant decrease in the pixel-level PSNR index. Although the LPIPS index improves continuously, excessive loss of fidelity performance might lead to the generation of unrealistic and visually over-enhanced details.  

In the right part of Fig. \ref{fig2}, we visualize the SR results of three LR images. When LR is corrupted heavily (the first row), SwinIR-GAN struggles to generate fine details based on the inadequate image structures restored by the $\ell_1$ loss, while StableSR produces more realistic results by exploiting the strong natural image priors. When sufficient structural information is available in the LR image (the second row), SwinIR-GAN can perform similarly well to StableSR to restore low-frequency structures, and both of them can reconstruct the HR image with visually pleasing details. However, due to the randomness in the synthesis process of DMs, the restored image structures and details are likely to be inconsistent with the LR input and the GT, even in the case of minor image degradation (bottom row). In contrast, SwinIR-GAN works well in terms of fidelity and content consistency for the bottom image.

To sum up, the DM-based methods showcase greater proficiency in learning complex natural image priors and refining image structures than GAN-based methods. Nonetheless, DM-based approaches encounter instability performance brought by the randomness introduced during the noise sampling process. On the other hand, GAN-based methods excel at augmenting deterministic details if the structure can be effectively reconstructed. However, GAN-based methods face challenges in restoring structures, making detail enhancement a formidable task for them. The performance discrepancy between DM-based and GAN-based methods enlarges as the LR quality deteriorates.


The above observations motivate us to propose a new framework to disentangle the SR process into structure generation and detail enhancement by GAN and DM, for a more stable and effective use of generative priors for SR. Our proposed framework, namely content-consistent super-resolution (CCSR), is shown in Fig. \ref{fig3}.
There are two training stages in CCSR, structure refinement (top left) and detail enhancement (top right). In the first stage, a non-uniform sampling strategy (bottom) is proposed, which applies a single timestep for information extraction from the LR input to improve stability and fidelity. Several more timesteps can be optionally employed for more image structure generation, and then the diffusion process is directly terminated. The output of the first stage is fed into the second stage, which aims to synthesize realistic details based on the structures reproduced in the first stage. Rather than employing an additional GAN network, we finetune the already existed VAE decoder with the adversarial loss so that it can perform feature decoding and detail enhancement simultaneously without introducing additional computational overhead. The two stages are detailed in the following sections.


\subsection{Structure Refinement Stage}

\noindent\textbf{Preliminaries.}
\label{sec:diffusion}
DM employs a forward process to gradually transform an input image $x_0$ into Gaussian noise $x_T \sim N(0, 1)$ in $T$ steps: 
$x_t = \sqrt{1 - \beta_t} \cdot x_{t-1} + \sqrt{\beta_t} \cdot \epsilon$, where $x_t$ is the noisy image at step $t$, $\beta_t$ controls the noise level, and $\epsilon$ is random noise of standard normal distribution. This process can be reformulated as:
\begin{equation}
x_t = \sqrt {{{\bar \alpha }_t}} \cdot x_{0} +  \sqrt{{1 - {{\bar \alpha }_t}}}  \cdot \epsilon,
\label{diffusion}
\end{equation}
where ${\alpha _t} = 1 - {\beta _t}$, and ${{\bar \alpha }_t} = \prod\nolimits_{i = 1}^t {{\alpha _i}}$.

The reverse process of DM iteratively recovers the original image $x_0$ sampled from $p\left( {{x_{t - 1}}\left| {{x_t},{x_0}} \right.} \right) = N\left( {{x_{t - 1}};\mu_t\left( {{x_t},{x_0}} \right),\sigma _t^2{\bf{I}}} \right)$. The mean of $x_{t-1}$ is ${\mu _t}\left( {{x_t},{x_0}} \right) = \frac {\sqrt {\overline \alpha_{t-1}} \beta_t}{1-\overline \alpha_t} x_0 + \frac {\sqrt {\alpha_t}(1-\overline \alpha_{t-1})}{1-\overline \alpha_t} x_t$, and the variance is $\sigma _t^2 = \frac{{1 - {{\bar \alpha }_{t - 1}}}}{{1 - {{\bar \alpha }_t}}}{\beta _t}$. DM typically applies a denoising network $\epsilon_\theta(x_t, t)$ to estimate the noise so that the original image details can be reconstructed. 
During DM training, the noisy image $x_t$ is generated by randomly selecting a timestep $t \in [0, T)$ and noise $\epsilon \sim N(0, 1)$ according to Eq. (\ref{diffusion}). The loss function $l_{diff}$ is:
\begin{equation}
{l_{diff}} = \left\| {\epsilon  - {\epsilon _\theta }\left( {\sqrt {{{\bar \alpha }_t}}  \cdot {x_0} + \sqrt{{1 - {{\bar \alpha }_t}}} \cdot \epsilon ,t} \right)} \right\|_2^2.
\label{loss}
\end{equation}

\begin{figure*}[ht]
	\centering 
	\includegraphics[scale=0.76]{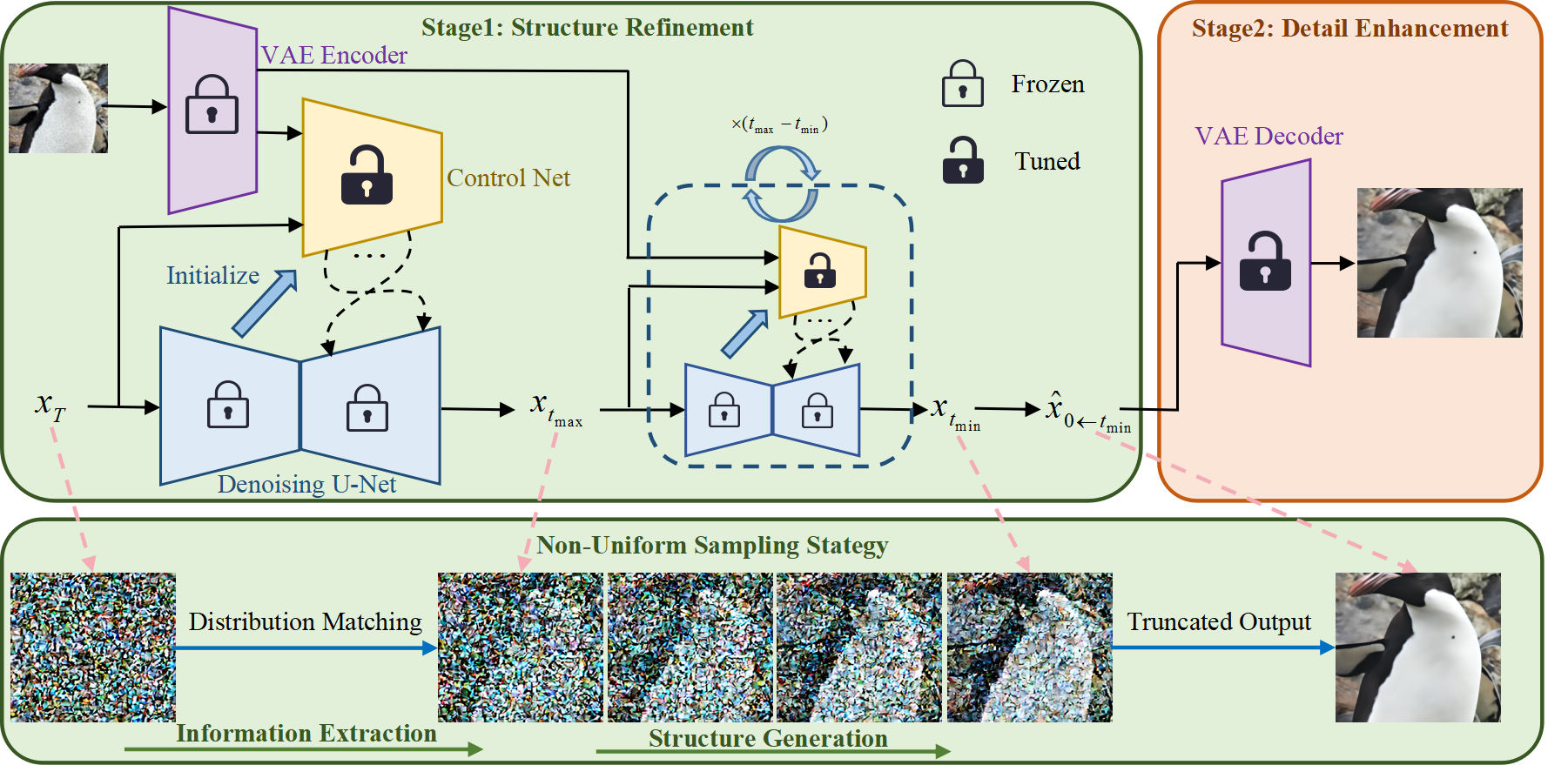}
	
	\caption{Framework of our proposed CCSR. There are two stages in CCSR, structure refinement (top left) and detail enhancement (top right). In the first stage, a non-uniform sampling strategy (bottom) is proposed, which applies one timestep for information extraction from LR and several other timesteps for image structure generation. The diffusion process is then stopped and the truncated output is fed into the second stage, where the detail is enhanced by finetuning the VAE decoder with adversarial training. }

\label{fig3}
\end{figure*}

\noindent\textbf{Non-Uniform Timestep Sampling.}
Most previous DM-based SR methods \cite{ldmsr, stablesr, resshift, diffbir, pasd, seesr, supir} follow the text-to-image generation methods \cite{controlnet} to employ a uniform sampling strategy with exhaustive iteration steps. However, text-to-image generation needs to generate almost every pixel from scratch, whereas in SR tasks an LR image is given, which provides the coarse structure for the desired image. 
The current noise sampling approaches do not fully take advantage of the LR input but iteratively generate the coarse structure, resulting in redundant computation, unwanted randomness, and losses in fidelity quality. As shown in Fig. \ref{fig1}, with the conventional uniform sampling strategy, the SR results of DM-based methods \cite{stablesr, pasd} with two random noise samples can be very different in textures and details.

As discussed in Sec. \ref{sec:motivation}, we propose to partition the diffusion process into two stages, as shown in Fig. \ref{fig3}. Note that the intermediate diffusion processes in Fig. \ref{fig3} are visualized by decoding. In the first stage of structure refinement, we propose a non-uniform sampling strategy to optimize the sampling process for SR tasks. For information extraction, only a single timestep sampling is required to extract the coarse information from the LR image by mapping the Gaussian noise $x_T$ to the intermediate noisy image $x_{t_{max}}$, which can guarantee the stability and fidelity in the diffusion process. For structure generation, the diffusion chain is truncated after a few uniform timesteps from $x_{t_{max}}$ to $x_{t_{min}}$. The truncated approach is used since the structure has already been well reconstructed in the intermediate process (please refer to the results of StableSR-600 in Fig. \ref{fig2}).
The estimated result from $x_{t_{min}}$, denoted as ${{\hat x}_{0 \leftarrow {t_{\min }}}}$, is output to the second stage of detail enhancement by generative adversarial training. 

Given $x_{t_{max}}$ and $x_{t_{min}}$, during the intervals $(T, t_{max})$ and $(t_{min}, 0]$, there is no need of sampling in training. The sampling is only needed when $t=T$ and $t$ falls into the range of $[t_{max}, t_{min}]$. 
The reverse process from $x_T$ to $x_{t_{max}}$ can be traced by substituting the corresponding parameter in Eq. (\ref{diffusion}). However, the diffusion step from $T$ to $t_{max}$ is much bigger than the original step so the Gaussian noise assumption will not hold \cite{truncateddiffusion, diffusiongan}. Therefore, directly applying this non-uniform sampling strategy will lead to significant performance loss. To solve this issue, we propose a non-uniform timestep sampling method with a newly designed training loss at $t=T$.

We propose to constrain the estimated ${{\hat x}_{0 \leftarrow T}}$ at $t=T$ rather than the sampled noise for extracting structure information from LR. Given a sampled start point $x_T$ by Eq. (\ref{diffusion}), the estimated noise $\hat \epsilon_T$ can be obtained from the denoising network by $\hat \epsilon_T = \epsilon_\theta(x_T, T)$. Then ${{\hat x}_{0 \leftarrow T}}$ can be calculated by ${{\hat x}_{0 \leftarrow T}} = \frac{1}{{\sqrt {{{\bar \alpha }_T}} }}\left( {{x_T} - \sqrt { {1 - {{\bar \alpha }_T}}} \cdot \hat \epsilon_T } \right)$. Consequently, the loss function for $t=T$ is $l_T=\left\| {{x_0} - {{\hat x}_{0 \leftarrow T}}} \right\|_2^2$. Using the estimated ${{\hat x}_{0 \leftarrow T}}$, $\hat x_{t_{max}}$ can be obtained by adding the corresponding noise as ${\hat x}_{t_{max}} = \sqrt {{{\bar \alpha }_{t_{max}}}} \cdot {{\hat x}_{0 \leftarrow T}} +  \sqrt{{1 - {{\bar \alpha }_{t_{max}}}}}  \cdot \epsilon$. To preserve the continuity of the diffusion chain, we enforce the same constraint on $\hat x_{t_{max}}$ as that on $x_T$, leading to ${l_{{t_{\max }}}} = \left\| {{x_0} - \frac{1}{{\sqrt {{{\bar \alpha }_{{t_{\max }}}}} }}\left( {{{\hat x}_{{t_{\max }}}} - \sqrt{ {1 - {{\bar \alpha }}_{{t_{\max }}}}} \hat \epsilon_{t_{max}} }\right) } \right\|_2^2$. In $l_{t_{max}}$, $\hat \epsilon_{t_{max}} = \epsilon_\theta(\hat x_{t_{max}}, t_{max})$.
Finally, the training loss of CCSR at $t=T$ is:
\begin{equation}
l_{diff}^T = {l_T} + {l_{{t_{\max }}}}.
\label{Tloss}
\end{equation}
Note that we do not change the loss function for the other sampling timesteps.

\subsection{Detail Enhancement Stage}

Based on the refined image structures in the first stage, we leverage adversarial training to enhance the fine details without introducing further randomness. While it is common to employ an additional module for enhancement \cite{resshift, dr2}, we adopt a more efficient approach by fine-tuning the already existed VAE decoder. This is motivated by previous findings \cite{zhu2023designing, dr2, luo2023image} that the VAE decoder has redundancy and untapped potential. In specific, we reuse the VAE decoder to decode latent features and enhance details. The training loss is the same as that of VAE \cite{vae}. Remarkably, this simple strategy achieves outstanding performance, as demonstrated in our ablation study in Sec. \ref{sec:Ablation} and in Fig. \ref{fig1}.

\subsection{Training Process}

We train Stage 1 of CCSR first and take its output from $x_{T}$ and ${{\hat x}_{0 \leftarrow {T}}}$ as the input of Stage 2. In the training of Stage 2, all parameters of the first stage are frozen. With our training strategy, the trained CCSR model can achieve different sampling steps for SR during inference. 
For multi-step diffusion, ${{\hat x}_{0 \leftarrow {t_{\min }}}}$ can be iteratively obtained from $x_{t_{max}}$ and it is set as the input of Stage 2. If ${{\hat x}_{0 \leftarrow {T}}}$ is directly set as the input of Stage 2, \textit{an efficient one-step diffusion model can be obtained}.

In the proposed CCSR framework, the overall diffusion reverse timesteps can be calculated by $S=(t_{max}-t_{min})*T+1$.
As the number of diffusion timesteps increases, the details of the reconstructed image become richer but the fidelity to input may decrease. In other words, increasing the number of diffusion timesteps improves the no-reference metrics of the restored images, but compromises their full-reference metrics.
When comparing with the multi-step DM-based SR methods, we set $T$, $t_{max}$, and $t_{min}$ as $6$, $\frac{{2}}{3}$, and $\frac{{1}}{2}$ in all our experiments.
The influence of different selections of $T$, $t_{max}$, and $t_{min}$ will be discussed in Sec. \ref{sec:Ablation}.


\section{EXPERIMENT}
\label{sec:Experiment}

\subsection{Experimental Setting}

\noindent\textbf{Training and Inference.}
CCSR is built upon ControlNet \cite{controlnet} with Stable Diffusion (SD) 2.1-base \cite{stablediffusion}. We first finetune the pre-trained SD for 25K iterations. In the adversarial training of the VAE decoder, we finetune it for 2K iterations. We use LSDIR \cite{lsdir} and the first 5K images in FFHQ \cite{ffhq} as the training data. The degradation pipeline in RealESRGAN \cite{realesrgan} is used to generate the paired training data for comparisons on real-world SR tasks. The Adam \cite{adam} optimizer is used in optimizing the models, and the learning rates of the two training stages are $5e^{-5}$ and $1e^{-5}$. The batch sizes in both the two training stages are set as 96 and 64. The size of training patches is 512×512. 

During inference, we use a spaced DDPM sampling method \cite{ddpm, spacedddpm} with our proposed non-uniform sampling strategy. We found that setting $T = 6, t_{max}=2/3, t_{min} = 1/2$ with \textit{two diffusion steps} is enough for our CCSR method to produce appealing visual and numerical results compared with other DM-based SR methods. Furthermore, we show that adopting only \textit{one diffusion step} in the CCSR framework can also achieve competitive results.

\noindent\textbf{Compared Methods.}  We compare CCSR with representative and state-of-the-art GAN-based methods, and standard and efficient DM-based SR methods. The GAN-based SR methods include BSRGAN \cite{bsrgan} and RealESRGAN \cite{realesrgan}. The standard DM-based SR methods include StableSR \cite{stablesr}, ResShift \cite{resshift}, DiffBIR \cite{diffbir}, PASD \cite{pasd}, SeeSR \cite{seesr} and SUPIR \cite{supir}, which run tens to hundreds of diffusion steps. The efficient DM-based SR methods include SinSR \cite{sinsr}, AddSR \cite{AddSR} and OSEDiff \cite{osediff}, which require less than 5 diffusion steps and even only one step diffusion.  
The results of the compared methods are obtained by using their officially released codes or models.  For fairness, we use the default diffusion timesteps of the competing DM-based methods. We also report the SR performance of standard DR-based methods (StableSR, ResShift, DiffBIR, PASD, SeeSR and SUPIR) with 3 diffusion steps to further show the advantage of our method. 
 
\noindent\textbf{Test Datasets.} To comprehensively evaluate the effectiveness of our CCSR method, we conduct experiments on the following real-world and synthetic datasets. 

\begin{itemize}
  \item The cropped RealSR \cite{realsr} and DRealSR \cite{drealsr} datasets released in \cite{stablesr}, where the images suffer from real-world unknown degradations. 
  \item The degraded DIV2K \cite{div2k} test set in \cite{stablesr} following the degradation pipeline of RealESRGAN \cite{realesrgan}. 
  
 
\end{itemize}

The LR images are cropped to $128 \times 128$, and resized to $512 \times 512$ as the input by the bicubic interpolation method, following StableSR \cite{stablesr}.

\subsection{Evaluation Metrics}

\noindent\textbf{Existing Quality Measures.} Following \cite{stablesr, LDL}, we use the following reference and no-reference metrics to compare the performance of different methods: 
\begin{itemize}
  \item PSNR and SSIM \cite{ssim}, computed on the Y channel in the YCbCr space, to measure the fidelity of SR results. 
  \item LPIPS \cite{lpips} and DISTS \cite{dists}, computed in the RGB space, to evaluate the perceptual quality of SR results.  
  \item No-reference image quality metrics NIQE, CLIPIQA \cite{clipiqa}, MUSIQ \cite{musiq} and MANIQA \cite{maniqa}.
  \item FID \cite{fid}, computed in the RGB space, to measure the statistical distance between real images and SR results using a pre-trained Inception network. 
\end{itemize}
It should be noted that for DM-based methods, each value of the above metrics is calculated by \textit{averaging the results over 10 runs with 10 different noise samples}.

\noindent\textbf{New Stability Measures.} As mentioned in Sec. \ref{sec:intro}, enhancing the stability of DM-based SR methods is vital to ensure that they can produce more reliable outputs. Considering that most existing DM-based SR techniques suffer from the stability problem, \ie, they may generate different results of various quality with different samples (see Fig. \ref{fig1} for example), it is necessary to design stability measures for a more comprehensive and fair comparison of the DM-based methods.

We make such an attempt in this paper and propose two stability metrics, namely global standard deviation (\textbf{G-STD}) and local standard deviation (\textbf{L-STD}), to measure the image-level and pixel-level variations of the SR results. We run $N$ times ($N=10$ in this paper) the experiments for each SR model on each test image within each test benchmark.
For each SR image, we can compute its quality metrics (except for FID) and then calculate the STD over the $N$ runs for each metric. By averaging the STD values over all test images in a benchmark, the G-STD value of one metric, denoted by $p$, can be obtained: 
\begin{equation}
\text{{G-{STD}}}^{p} = \frac{1}{{M}}{\sum_{j=1}^{M}}\sqrt{\frac{{\sum_{i=1}^{N} (p_i^j - \bar{p}^j)^2}}{N}},
\label{gstd}
\end{equation}
where $p_{i}^j$ is the value of $p$ for the restored image in the $i$-th run for the $j$-th image in a dataset with $M$ images, and $\bar{p}^j$ is the average of $p^j$ over $N$ runs. 

G-STD reflects the stability of an SR model at the image level. To measure the stability at the local pixel level, we define L-STD, which computes the STD of pixels in the same location of the $N$ SR images: 
\begin{equation}\small
\text{{L-{STD}}}=\frac{1}{{MHW}}\sum_{j = 1}^M\sum_{h = 1}^H {\sum_{w = 1}^W {\sqrt {\frac{{\sum_{i = 1}^N {\left( {x_{i,(h,w)}^j - {{\bar x}_{(h,w)}^j}} \right)} }}{N}} } },
\label{lstd}
\end{equation}
where $x_i^{j}$ denotes the restored image in the $i$-th run for the $j$-th image in a dataset, $H$ and $W$ denote image height and weight, $(h,w)$ denote pixel location, and $\bar{x}_{(h,w)}$ is the mean of the $N$ pixels at $(h,w)$.

\subsection{Ablation Studies}
\label{sec:Ablation}
In this section, we first perform ablation studies to validate the effectiveness of our proposed non-uniform timestep sampling (NUTS) and VAE decoder finetuning (DeFT) strategies, and then discuss the selection of $T$, $t_{max}$, and $t_{min}$, which  determine the number of diffusion steps, \ie, `S'.

\begin{figure*}
	\centering 
	\includegraphics[scale=1.15]{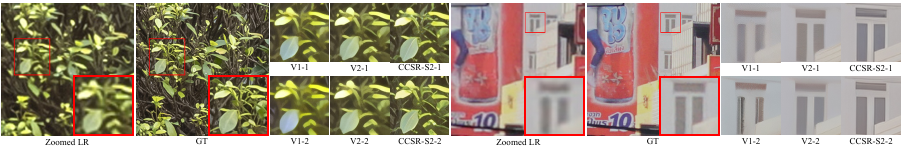}
	
	\caption{Visual comparisons of CCSR and its variants `V1' and `V2'. One can see that the NUTS and DeFT strategies improve the super-resolution performance and stability. }
	\label{fig: ablation}
 
\end{figure*}

\noindent\textbf{The Effectiveness of NUTS and DeFT.} Table \ref{ablation} and Fig. \ref{fig: ablation} validate the effectiveness of NUTS and DeFT strategies. We define two variants of CCSR, \ie, removing both the NUTS and DeFT strategies (see `V1') and removing only DeFT (see `V2').
As can be seen in Fig. \ref{fig: ablation}, the results of  `V1' exhibit noticeable color distortion and disorganized details. This phenomenon stems from the ineffective utilization of information in the LR input when only two diffusion steps are applied for restoration. By introducing the NUTS strategy, the variant `V2' improves all metrics, as shown in  Table \ref{ablation}, and reduces the visual artifacts, as shown in Fig. \ref{fig: ablation}. Finally, by integrating both NUTS and DeFT into CCSR, most of the perception metrics can be further improved, and the restored images showcase the best visual quality. Note that the introduction of DeFT slightly amplifies the variation of the output in the first stage of CCSR so that the G-STD and L-STD of CCSR-S2 are a little higher than` V2'.

\begin{table*}[]
\renewcommand{\arraystretch}{1.2}
\caption{
\begin{justify}
        \justifying 
Ablation studies on the proposed non-uniform timestep sampling (NUTS) and VAE decoder finetuning (DeFT) strategies on RealSR \cite{realsr} and DrealSR \cite{drealsr} benchmarks. We implement two variants of CCSR. `V1' means removing both NUTS and DeFT strategies, and `V2' means removing the DeFT strategy only. 
\end{justify}
}
\label{ablation}
\centering
\resizebox{\linewidth}{!}{
\begin{tabular}{c|ccc|ccccccc}
\hline
Datasets                 & Methods & NUTS & DeFT & PSNR/G-STD& LPIPS/G-STD& DISTS/G-STD& CLIPIQA/G-STD& MUSIQ/G-STD  & MANIQA/G-STD  & L-STD  \\ \hline
\multirow{3}{*}{RealSR}  & V1      &      $\color{red}{\times}$&      $\color{red}{\times}$& 25.71/0.3229                                          & 0.3634/0.0175                                          & 0.2900/0.0096                                          & 0.5898/0.0489                                            & 60.88/2.5251 & 0.5042/0.0259 & 0.0194 \\
                         & V2      &      \color{blue}{$\checkmark$}&      $\color{red}{\times}$& 26.71/0.2236                                          & 0.3172/0.0083                                          & 0.2667/0.0073                                          & 0.6166/0.0467                                            & 62.85/2.0983 & 0.5391/0.0244 & 0.0142 \\
                         & CCSR-S2&      \color{blue}{$\checkmark$}&      \color{blue}{$\checkmark$}& 25.86/0.2916                                          & 0.2941/0.0127                                          & 0.2296/0.0090                                          & 0.6561/0.0325                                            & 71.17/1.2133 & 0.6656/0.0140 & 0.0194 \\ \hline
\multirow{3}{*}{DrealSR} & V1      &      $\color{red}{\times}$&      $\color{red}{\times}$& 28.85/0.4185                                          & 0.3648/0.0218                                          & 0.3035/0.0121                                          & 0.5481/0.0549                                            & 54.44/3.1238 & 0.4311/0.0277 & 0.0165 \\
                         & V2      &      \color{blue}{$\checkmark$}&      $\color{red}{\times}$& 29.86/0.2871                                          & 0.3232/0.0100                                          & 0.2766/0.0088                                          & 0.5931/0.0533                                            & 56.30/2.6213 & 0.4626/0.0283 & 0.0120 \\
                         & CCSR-S2&      \color{blue}{$\checkmark$}&      \color{blue}{$\checkmark$}& 28.43/0.4366                                          & 0.3397/0.0181                                          & 0.2563/0.0125                                          & 0.6695/0.0299                                            & 68.49/1.4207 & 0.6332/0.0173 & 0.0183 \\ \hline
\end{tabular}
}
\end{table*}
\begin{table*}
\renewcommand{\arraystretch}{1.2}
\caption{
\begin{justify}
Ablation studies on the selection of  $T$ when keeping $t_{max}=2/3$ and $t_{min} = 1/2$ on RealSR \cite{realsr} and DrealSR \cite{drealsr} benchmarks. $S$ denotes the number of diffusion steps.
\end{justify}}

\label{ablation_T}
\centering
\resizebox{\linewidth}{!}{
\begin{tabular}{c|c|cccccccccc}
\hline
Datasets                 & (T,S)  & PSNR/G-STD& SSIM/G-STD& LPIPS/G-STD& DISTS/G-STD& FID    & NIQE/G-STD& CLIPIQA/G-STD& MUSIQ/G-STD  & MANIQA/G-STD  & L-STD  \\ \hline
\multirow{3}{*}{RealSR}  & (18,4) & 25.45/0.3257                                          & 0.7168/0.0142                                         & 0.3063/0.0154                                          & 0.2347/0.0096                                          & 129.95 & 5.99/0.4725                                           & 0.6671/0.0334                                            & 72.13/1.0324 & 0.6762/0.0135 & 0.0232 \\
                         & (12,3) & 25.59/0.3081                                          & 0.7227/0.0127                                         & 0.3025/0.0135                                          & 0.2335/0.0094                                          & 129.25 & 6.02/ 0.4626                                          & 0.6669/0.0329                                            & 71.91/1.0034 & 0.6742/0.0135 & 0.0220 \\
                         & (6,2)  & 25.86/0.2916                                          & 0.7335/0.0115                                         & 0.2941/0.0127                                          & 0.2296/0.0090                                          & 126.32 & 6.07/0.4632                                           & 0.6561/0.0325                                            & 71.17/1.2133 & 0.6656/0.0140 & 0.0194 \\ \hline
\multirow{3}{*}{DrealSR} & (18,4) & 28.07/0.4515                                          & 0.7563/0.0196                                         & 0.3552/0.0214                                          & 0.2625/0.0118                                          & 169.41 & 6.85/0.6950                                           & 0.6882/0.0293                                            & 69.51/1.3556 & 0.6441/0.0171 & 0.0213 \\
                         & (12,3) & 28.21/0.4402                                          & 0.7626/0.0180                                         & 0.3482/0.0196                                          & 0.2600/0.0126                                          & 167.36 & 6.95/0.6772                                           & 0.6844/0.0301                                            & 69.12/1.3866 & 0.6419/0.0173 & 0.0202 \\
                         & (6,2)  & 28.43/0.4366                                          & 0.7724/0.0172                                         & 0.3397/0.0181                                          & 0.2563/0.0125                                          & 163.74 & 7.00/0.6728                                           & 0.6695/0.0299                                            & 68.49/1.4207 & 0.6332/0.0173 & 0.0183 \\ \hline
\end{tabular}
}
\end{table*}

\begin{table*}
\renewcommand{\arraystretch}{1.2}
\caption{
\begin{justify}
Ablation studies on the selection of  $t_{max}$  and $t_{min}$ when keeping $T=6$ on RealSR \cite{realsr} and DrealSR \cite{drealsr} benchmarks. $S$ denotes the number of diffusion steps.
\end{justify}
}

\label{ablation_t_maxmin}
\centering
\resizebox{\linewidth}{!}{
\begin{tabular}{c|c|cccccccccc}
\hline
Datasets                 & ($t_{max}$,$t_{min}$,S)& PSNR/G-STD& SSIM/G-STD& LPIPS/G-STD& DISTS/G-STD& FID    & NIQE/G-STD& CLIPIQA/G-STD& MUSIQ/G-STD  & MANIQA/G-STD  & L-STD  \\ \hline
\multirow{5}{*}{RealSR}  & (2/3,1/3,3)       & 25.90/0.3032                                          & 0.7339/0.0115                                         & 0.2942/0.0135                                          & 0.2288/0.0096                                          & 128.88 & 6.16/0.4838                                           & 0.6626/0.0347                                            & 70.82/1.3055 & 0.6618/0.0156 & 0.0220 \\
                         & (5/6,1/2,3)       & 25.47/0.2862                                          & 0.7163/0.0121                                         & 0.3075/0.0131                                          & 0.2354/0.0089                                          & 130.02 & 6.00/0.4310                                           & 0.6748/0.0281                                            & 72.19/0.8895 & 0.6784/0.0131 & 0.0224 \\
                         & (5/6,2/3,2)       & 25.43/0.2659                                          & 0.7206/0.0115                                         & 0.3061/0.0115                                          & 0.2340/0.0078                                          & 130.87 & 5.85/0.4066                                           & 0.6657/0.0271                                            & 71.98/0.7616 & 0.6746/0.0111 & 0.0197 \\
                         & (1/2,1/6,2)       & 26.29/0.2594                                          & 0.7484/0.0089                                         & 0.2860/0.0115                                          & 0.2235/0.0090                                          & 126.47 & 6.39/0.0090                                           & 0.6146/0.0379                                            & 68.40/1.6272 & 0.6319/0.0160 & 0.0175 \\
                         & (2/3,1/2,2)& 25.86/0.2916                                          & 0.7335/0.0115                                         & 0.2941/0.0127                                          & 0.2296/0.0090                                          & 126.32 & 6.07/0.4632                                           & 0.6561/0.0325                                            & 71.17/1.2133 & 0.6656/0.0140 & 0.0194 \\ \hline
\multirow{5}{*}{DrealSR} & (2/3,1/3,3)       & 28.49/0.4128                                          & 0.7707/0.0169                                         & 0.3416/0.0195                                          & 0.2581/0.0123                                          & 162.82 & 7.04/0.7095                                           & 0.6695/0.0343                                            & 68.00/1.5719 & 0.6298/0.0200 & 0.0204 \\
                         & (5/6,1/2,3)       & 28.09/0.4383                                          & 0.7558/0.0184                                         & 0.3544/0.0202                                          & 0.2609/0.0122                                          & 166.71 & 6.95/0.7235                                           & 0.6888/0.0295                                            & 69.39/1.3386 & 0.6439/0.0162 & 0.0204 \\
                         & (5/6,2/3,2)       & 28.02/0.4018                                          & 0.7607/0.0167                                         & 0.3484/0.0166                                          & 0.2547/0.0101                                          & 164.43 & 6.91/0.6434                                           & 0.6884/0.0262                                            & 68.95/1.2711 & 0.6423/0.0138 & 0.0182 \\
                         & (1/2,1/6,2)       & 28.78/0.4189                                          & 0.7835/0.0160                                         & 0.3339/0.0181                                          & 0.2549/0.0116                                          & 164.25 & 7.32/0.0116                                           & 0.6126/0.0367                                            & 65.75/1.8625 & 0.5871/0.0208 & 0.0168 \\
                         & (2/3,1/2,2)& 28.43/0.4366                                          & 0.7724/0.0172                                         & 0.3397/0.0181                                          & 0.2563/0.0125                                          & 163.74 & 7.00/0.6728                                           & 0.6695/0.0299                                            & 68.49/1.4207 & 0.6332/0.0173 & 0.0183 \\ \hline
\end{tabular}
}
\end{table*}

\noindent\textbf{The Selection of $T$, $t_{max}$, and $t_{min}$.} 
Recall that in the proposed CCSR framework, the number of diffusion steps is determined as $S=(t_{max}-t_{min})*T+1$.
Table \ref{ablation_T}  shows the performance of CCSR with different $T$ by keeping $t_{max}=2/3$ and $t_{min}=1/2$. One can see that the full reference based metrics get worse but the no-reference based metrics get better with the increase of $T$. This is because with the increase of diffusion steps, more details will be generated but the fidelity will be reduced. 
Table \ref{ablation_t_maxmin} tests several alternative selections of $t_{max}$ and $t_{min}$, \ie, $t_{max}, t_{min} = \frac{{2}}{3}, \frac{{1}}{3}$, $t_{max}, t_{min} = \frac{{5}}{6}, \frac{{1}}{2}$, $t_{max}, t_{min} = \frac{{5}}{6}, \frac{{2}}{3}$, and $t_{max}, t_{min} = \frac{{1}}{2}, \frac{{1}}{6}$ with $T=6$. The numbers of diffusion steps are 3, 3, 2, and 2 accordingly. 

We can see that by keeping $t_{max}$ as a constant, using a larger $t_{min}$ can achieve better results in both reference-based and no-reference metrics with fewer diffusion timesteps. This indicates that the latter part of the diffusion process has a counterproductive effect on the SR output, further validating the effectiveness of our truncated strategy in the first stage of CCSR. When $t_{min}$ is kept constant, setting a larger $t_{max}$ can achieve higher no-reference metrics. However, those reference-based metrics and stability metrics deteriorate simultaneously. This suggests that using more diffusion steps in the early stage will destroy the LR structure information.
When $S$ is kept constant, using a larger interval between  $t_{max}$ and  $t_{min}$ in the early stage could improve the no-reference metrics. Conversely, a larger interval between them in the later stage could improve the reference-based metrics. Overall, the perception-fidelity trade-off can be achieved by adjusting $T$, $t_{max}$, and $t_{min}$. 

In all our following experiments, we set $T = 6, t_{max}=2/3, t_{min} = 1/2$ for CCSR with 2 diffusion steps (\ie, CCSR-S2). In addition, we also verify the performance of CCSR with only 1 diffusion step (\ie, CCSR-S1) by directly setting ${{\hat x}_{0 \leftarrow {T}}}$ as the input of the VAE decoder.

\subsection{Comparisons with Standard DM-based SR Methods}
\noindent\textbf{Quantitative Comparisons.} We first perform comparison with those standard DM-based SR methods, which cost tens to hundreds of diffusion steps. These methods can be divided into two types. The first type uses an adapter \cite{controlnet} to finetune a pre-trained text-to-image diffusion model, including StableSR \cite{stablesr}, DiffBIR \cite{diffbir}, PASD \cite{pasd}, SeeSR \cite{seesr}, SUPIR \cite{supir} and our CCSR. Another type trains a model from scratch, \ie, ResShift \cite{resshift}, which redefines a diffusion reverse process for the SR task and shows rather different behaviors from other DM-based methods. The results are shown in Table \ref{comparison-1}. We can have the following observations. 

First, there are notable distinctions between GAN-based and DM-based SR methods. Due to the stronger generation capability, most DM-based methods perform better in no-reference indices, such as NIQE, CLIPIQA, and MUSIQ, while sacrificing fidelity performance. For example, SeeSR \cite{seesr} outperforms BSRGAN by 5.6 in MUSIQ, while its PSNR is 1dB lower on the RealSR dataset. 
ResShift \cite{resshift} uses a redefined diffusion chain to train the DM from scratch, achieving better fidelity indices but lower perceptual quality (see the sub-section of qualitative comparison).

Second, the existing DM-based methods can only achieve optimal performance in either fidelity quality or perceptual quality, and CCSR performs favorably against other methods in fidelity and perceptual-related measures. In terms of fidelity metrics (PSNR and SSIM), CCSR-S1 and ResShift perform similarly, and CCSR-S2 is only slightly worse than ResShift. However, both the two CCSR models achieve significantly better perceptual metrics with improved visual quality. For both full-reference perceptual metrics (LPIPS and DISTS) and no-reference ones (CLIPIQA, MUSIQ, MANIQA), CCSR achieves the most competitive results, with the best or second-best results in almost all metrics among all DM-based SR methods across all the test sets. In particular, CCSR-S2 obtains the best MUSIQ score in all the test sets, although it only uses two steps.

\begin{figure*}
	\centering 
	\includegraphics[scale=1.28]{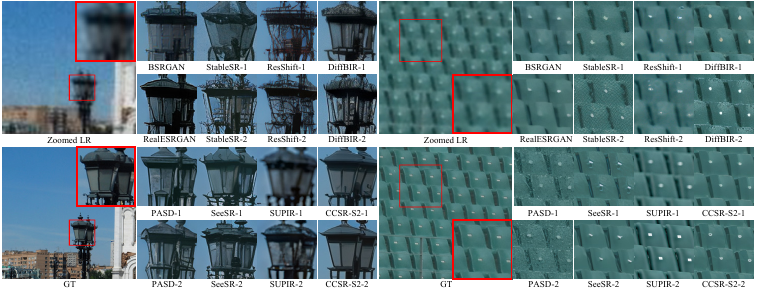}
 
	\caption{Visual comparisons (better zoom-in on screen) between CCSR and state-of-the-art GAN-based and the standard DM-based SR methods. For each of the DM-based methods, two restored images that have the best and worst PSNR values over $10$ runs are shown for a more comprehensive and fair comparison. Our proposed CCSR works the best to reconstruct accurate structures and realistic, content-consistent and stable details.}

	\label{fig5}
\end{figure*}

Last but not least, as a DM-based SR method, CCSR demonstrates much better stability in synthesizing image details, as evidenced by its outstanding G-STD and L-STD measures. 
Specifically, CCSR achieves the best L-STD scores on all the test sets, showcasing its strong capability in reducing the stochasticity of local structure and detail generation. 
It achieves most of the best G-STD scores on reference metrics, and the best and second-best G-STD scores on no-reference metrics, demonstrating high content consistency of SR outputs. Though ResShift also has good stability measures, its visual quality is less satisfactory (see Fig. \ref{fig5}).

\noindent\textbf{Qualitative Comparisons.}
We present visual comparisons in Fig. \ref{fig5}. Considering the stochasticity in DMs, the restored images with the best and worst PSNR values over $10$ runs are given for each DM-based SR method for a more fair comparison. 
One can see that GAN-based methods are difficult to generate textures from the degraded structures in the LR image, resulting in over-smoothed or even wrong details (\eg, the streetlight in the left group). 
Among the DM-based methods, ResShift has relatively lower perceptual quality, failing to synthesize realistic structures (\eg, bleacher seat in the right group).
StableSR, DiffBIR, PASD, SeeSR, and SUPIR can generate perceptually more realistic details by leveraging the strong diffusion priors in the pre-trained SD model; however, their outputs are unstable. The two results with the highest and lowest PSNR values can vary a lot. In contrast, our proposed CCSR can produce high-quality realistic SR results and have high stability. One can see that the two images with the best and worst PSNR values produced by CCSR only vary a little in content. 

\begin{table*}
           \centering

 \caption{
 \begin{justify}
        \justifying
        Quantitative comparison among the state-of-the-art GAN-based SR methods and standard DM-based SR methods, which require tens to hundreds of diffusion steps, on both synthetic and real-world test datasets. $S$ denotes the number of diffusion steps. Note that the G-STD is not available for FID, because FID measures the statistical distance between two groups of images. The best and the second-best results are highlighted in {\color[HTML]{FF0000} \textbf{red}} and  {\color[HTML]{2E75B5} \textbf{blue}}, respectively.
        \end{justify}}
 
   \resizebox{\linewidth}{!}{
\begin{tabular}{c|c|cccccccccc}
\hline
Datasets                   & Methods                                                  & PSNR/G-STD&SSIM /G-STD& LPIPS /G-STD& DISTS /G-STD& FID                                    & NIQE /G-STD& CLIPIQA /G-STD& MUSIQ/G-STD                                  & MANIQA/G-STD                                  & L-STD                                  \\ \hline
                           & BSRGAN                                                   & {\color[HTML]{2E75B5} \textbf{24.60}}/-                & 0.6268/-                                                 & 0.3361/-                                                   & 0.2268/-                                                   & 44.22                                  & 4.75/-                                                    & 0.5204/-                                                     & 61.16/-                                          &                                               0.5071/- & -                                      \\
                           & Real-ESRGAN                                              & 24.33/-                                                 & {\color[HTML]{FF0000} \textbf{0.6372}}/-                 & {\color[HTML]{FF0000} \textbf{0.3124}}/-                 & 0.2135/-                                                 & 37.64                                  & 4.68/-                                                  & 0.5219/-                                                   & 60.92/-                                        &      0.5501/-  & -                                      \\
                           & ResShift-S15                                             & {\color[HTML]{FF0000} \textbf{24.69}}/0.2720          & 0.6175/0.0118                                         & 0.3374/0.0196                                          & 0.2215/0.0116                                          & 36.01                                  & 6.82/0.5025                                           & 0.6089/0.0537                                            & 60.92/2.7917                                 &                                               0.5450/0.0200& 0.0340                                 \\
                           & StableSR-S200                                            & 23.31/0.4874                                          & 0.5728/0.0250                                         & 0.3129/0.0303                                          & 0.2138/0.0166                                          & {\color[HTML]{FF0000} \textbf{24.67}}  & 4.76/0.5673                                           & 0.6682/0.0592                                            & 65.63/3.4023                                 &                                               0.6188/0.0259& 0.0411                                 \\
                           & DiffBIR-S50                                              & \multicolumn{1}{r}{23.67/0.6910}                      & 0.5653/0.0396                                         & 0.3541/0.0466                                          & 0.2129/0.0220                                          & 30.93                                  & {\color[HTML]{0070C0} \textbf{4.71}}/0.7515           & 0.6652/0.0817                                            & 65.66/4.3691                                 &                                               0.6204/0.0339& 0.0443                                 \\
                           & PASD-S20                                                 & \multicolumn{1}{r}{23.14/0.5489}                      & 0.5489/0.0248                                         & 0.3607/0.0311                                          & 0.2219/0.0142                                          & 29.32                                  & {\color[HTML]{FF0000} \textbf{4.40}}/0.5747           & 0.6711/0.0442                                            & 68.83/2.2256                                 &                                               {\color[HTML]{FF0000} \textbf{0.6484}}/0.0239& 0.0430                                 \\
                           & SeeSR-S50                                                & 23.71/0.3921                                          & 0.6045/0.0143                                         & 0.3207/0.0196                                          & {\color[HTML]{FF0000} \textbf{0.1967}}/0.0121          & {\color[HTML]{0070C0} \textbf{25.83}}  & 4.82/0.5115                                           & {\color[HTML]{0070C0} \textbf{0.6857}}/0.0521           & 68.49/2.3691                                 & 0.6239/0.0245                                 & 0.0365                                 \\
                           & SUPIR-S50                                                & 23.57/0.3685                                          & 0.5665/0.0163                                         & 0.3819/0.0229                                          & 0.2310/0.0120                                      & 28.40                                  & 6.57/0.6072                                           & 0.6728/0.0408                                            & 59.69/2.9341                                 &                                               0.5635/0.0268& 0.0369                                 \\
                           & CCSR-S2
& 24.17/{\color[HTML]{0070C0} \textbf{0.2162}}                                          & 0.6130/ {\color[HTML]{0070C0} \textbf{0.0106 }}                                        & 0.3152/{\color[HTML]{0070C0} \textbf{0.0138}}                                          & 0.2216/{\color[HTML]{0070C0} \textbf{0.0102    }   }                                   & 36.08                                  & 5.62/{\color[HTML]{0070C0} \textbf{0.3798}}                                           & {\color[HTML]{FF0000} \textbf{0.7000}}/{\color[HTML]{0070C0} \textbf{0.0378}}            & {\color[HTML]{FF0000} \textbf{71.65}}/{\color[HTML]{FF0000} \textbf{1.1809}} & {\color[HTML]{0070C0} \textbf{0.6480/0.0154}} & {\color[HTML]{0070C0} \textbf{0.0265}} \\
\multirow{-10}{*}{DIV2K}   & CCSR-S1& 24.31/{\color[HTML]{FF0000} \textbf{0.1932}}          & {\color[HTML]{0070C0} \textbf{0.6283}}/{\color[HTML]{FF0000} \textbf{0.0082}}         & {\color[HTML]{FF0000} \textbf{0.2979}}/\color[HTML]{FF0000} \textbf{0.0111}         & {\color[HTML]{0070C0} \textbf{0.2020}}/{\color[HTML]{FF0000} \textbf{0.0083}}          & 30.83                                  & 5.32/{\color[HTML]{FF0000} \textbf{0.2982}}                                           & 0.6754/{\color[HTML]{FF0000}\textbf{0.0298}}                                            & {\color[HTML]{0070C0} \textbf{69.52}}/{\color[HTML]{0070C0} \textbf{1.1905}} & 0.6187/{\color[HTML]{FF0000} \textbf{0.0136}}& {\color[HTML]{FF0000} \textbf{0.0201}} \\ \hline
                           & BSRGAN                                                   & {\color[HTML]{FF0000} \textbf{26.39}}/-                 & {\color[HTML]{FF0000} \textbf{0.7654}}/-                & {\color[HTML]{FF0000} \textbf{0.2670}}/-                 & {\color[HTML]{0070C0} \textbf{0.2121}}/-                 & 141.28                                 & 5.66/-                                                  & 0.5001/-                                                  & 63.21/-                                        &                 0.5399/- & -                                      \\
                           & Real-ESRGAN                                              & 25.69/-                                                  & {\color[HTML]{0070C0} \textbf{0.7616}}/-                 & {\color[HTML]{0070C0} \textbf{0.2727}}/-                  & {\color[HTML]{FF0000} \textbf{0.2063}}/-                  & 135.18                                 & 5.83/-                                                   & 0.4449/-                                                    & 60.18/-                                         &                                               0.5487/- &                                        -\\
                           & ResShift-S15                                             & {\color[HTML]{0070C0} \textbf{26.31/0.2859}}          & 0.7411/0.0133                                         & 0.3489/0.0236                                          & 0.2498/0.0093                                          & 142.81                                 & 7.27/0.5592                                           & 0.5450/0.0493                                            & 58.10/2.5458                                 &                                               0.5305/0.0204& 0.0240                                 \\
                           & StableSR-S200                                            & 24.69/0.5600                                          & 0.7052/0.0219                                         & 0.3091/0.0299                                          & 0.2167/0.0152                                          & 127.20                                 & 5.76/0.6691                                           & 0.6195/0.0575                                            & 65.42/3.1678                                 &                                               0.6211/0.0251& 0.0300                                 \\
                           & DiffBIR-S50                                              & 24.88/0.7956                                          & 0.6673/0.0462                                         & 0.3567/0.0562                                          & 0.2290/0.0225                                          & 124.56                                 & 5.63/1.0350                                           & 0.6412/0.0739                                            & 64.66/4.6444                                 &                                               0.6231/0.0346& 0.0346                                 \\
                           & PASD-S20                                                 & 25.22/0.5301                                          & 0.6809/0.0275                                         & 0.3392/0.0311                                          & 0.2259/0.0130                                          & 123.08                                 & {\color[HTML]{FF0000} \textbf{5.18}}/0.6650           & 0.6502/0.0411                                            & 68.74/2.1633                                 &                                               {\color[HTML]{0070C0} \textbf{0.6461}}/0.0218& 0.0304                                 \\
                           & SeeSR-S50                                                & 25.33/0.4573                                          & 0.7273/0.0161                                         & 0.2985/0.0185                                          & 0.2213/0.0115                                          & 125.66                                 & {\color[HTML]{0070C0} \textbf{5.38}}/0.5242           & {\color[HTML]{FF0000} \textbf{0.6594}}/0.0510            & {\color[HTML]{0070C0} \textbf{69.37}}/1.7834 & 0.6439/0.0206& 0.0255                                 \\
                           & SUPIR-S50                                                & 25.20/0.5047                                          & 0.6916/0.0215                                         & 0.3582/0.0257                                          & 0.2423/0.0121                                          & {\color[HTML]{0070C0} \textbf{123.31}} & 7.18/0.6978                                           & 0.6371/0.0446                                            & 60.17/2.7544                                 &                                               0.5712/0.0228& 0.0253                                 \\
                           & CCSR-S2& 25.86/0.3032                                          & 0.7335/{\color[HTML]{0070C0} \textbf{0.0115}}& {\color[HTML]{0070C0} \textbf{0.2941}}/{\color[HTML]{0070C0} \textbf{0.0135}}& 0.2295/{\color[HTML]{0070C0} \textbf{0.0096}}& 126.12                                 & 6.07/{\color[HTML]{0070C0} \textbf{0.4838}}                                           & {\color[HTML]{0070C0} \textbf{0.6561}}/{\color[HTML]{0070C0} \textbf{0.0347}}            & {\color[HTML]{FF0000} \textbf{71.17}}/{\color[HTML]{0070C0} \textbf{1.3055}} & {\color[HTML]{FF0000} \textbf{0.6656}}/{\color[HTML]{0070C0} \textbf{0.0156}} & {\color[HTML]{0070C0} \textbf{0.0194}} \\
\multirow{-10}{*}{RealSR}  & CCSR-S1& 25.97/{\color[HTML]{FF0000} \textbf{0.1976}}& 0.7493/{\color[HTML]{FF0000} \textbf{0.0070}}& 0.2804/{\color[HTML]{FF0000} \textbf{0.0077}}& {\color[HTML]{0070C0} \textbf{0.2121}}/{\color[HTML]{FF0000} \textbf{0.0058}}& {\color[HTML]{FF0000} \textbf{121.43}} & 5.80/{\color[HTML]{FF0000} \textbf{0.3474 }}                                          & 0.6278/{\color[HTML]{FF0000} \textbf{0.0256}}                                            & 69.17/{\color[HTML]{FF0000} \textbf{0.9194 }}                                & 0.6405/{\color[HTML]{FF0000} \textbf{0.0105}}                                 & {\color[HTML]{FF0000} \textbf{0.0140}} \\ \hline
                           & BSRGAN                                                   & {\color[HTML]{FF0000} \textbf{28.75}}/-                  & {\color[HTML]{0070C0} \textbf{0.8031}}/-                 & {\color[HTML]{0070C0} \textbf{0.2883}}/-                  & {\color[HTML]{0070C0} \textbf{0.2142}}/-                                                  & 155.63                                 & 6.52/-                                                   & 0.4915/-                                                    & 57.14/-                                         &                                              0.4878/-&                                        -\\
                           & Real-ESRGAN                                              & {\color[HTML]{0070C0} \textbf{28.64}}/-                 & {\color[HTML]{FF0000} \textbf{0.8053}}/-                & {\color[HTML]{FF0000} \textbf{0.2847}}/-                 & {\color[HTML]{FF0000} \textbf{0.2089}}/-                & {\color[HTML]{0070C0} \textbf{147.62}} & 6.69/-                                                  & 0.4422/-                                                   & 54.18/-                                        &                                               0.4907/-   &                                        -\\
                           & ResShift-S15                                             & 28.45/{\color[HTML]{0070C0} \textbf{0.4100}}                                          & 0.7632/0.0197                                         & 0.4073/0.0349                                          & 0.2700/0.0132                                          & 175.92                                 & 8.28/0.5985                                           & 0.5259/0.0558                                            & 49.86/3.5063                                 &                                               0.4573/0.0279& 0.0241                                 \\
                           & StableSR-S200                                            & 28.04/0.7488                                          & 0.7460/0.0318                                         & 0.3354/0.0408                                          & 0.2287/0.0190                                          & {\color[HTML]{FF0000} \textbf{147.03}} & 6.51/0.8212                                           & 0.6171/0.0685                                            & 58.50/4.6598                                 &                                               0.5602/0.0351& 0.0257                                 \\
                           & DiffBIR-S50                                              & 26.84/1.3261                                          & 0.6660/0.0779                                         & 0.4446/0.0785                                          & 0.2706/0.0328                                          & 167.38                                 & {\color[HTML]{0070C0} \textbf{6.02}}/1.1834           & 0.6292/0.0904                                            & 60.68/6.1450                                 &                                               0.5902/0.0457& 0.0349                                 \\
                           & PASD-S20                                                 & 27.48/0.6497                                          & 0.7051/0.0304                                         & 0.3854/0.0333                                          & 0.2535/0.0147                                          & 157.36                                 & {\color[HTML]{FF0000} \textbf{5.57}}/0.7560& {\color[HTML]{FF0000} \textbf{0.6714}}/0.0467& 64.55/2.7189                                 &                                               {\color[HTML]{0070C0} \textbf{0.6130}}/0.0275& 0.0289                                 \\
                           & SeeSR-S50                                                & 28.26/0.6307                                          & 0.7698/0.0184                                         & 0.3197/0.0211                                          & 0.2306/0.0136                                          & 149.86                                 & 6.52/0.7485                                           & 0.6672/0.0491                                            & 64.84/2.8756                                 & 0.6026/0.0283                                 & 0.0229                                 \\
                           & SUPIR-S50                                                & 27.44/0.7986                                          & 0.6961/0.0409                                         & 0.4217/0.0419                                          & 0.2737/0.0149                                          & 153.35                                 & 9.43/1.1342                                           & 0.6035/0.0487                                            & 51.88/3.7709                                 &                                               0.5048/0.0326& 0.0263                                 \\
                           & CCSR-S2
& 28.44/0.4365                                     & 0.7724/{\color[HTML]{0070C0} \textbf{0.0172}}                                         & 0.3397/{\color[HTML]{0070C0} \textbf{0.0181}}                                          & 0.2563/{\color[HTML]{0070C0} \textbf{0.0123}}                                          & 161.94                                 & 7.01/{\color[HTML]{0070C0} \textbf{0.6728}}& {\color[HTML]{0070C0} \textbf{0.6695/0.0343}}            & {\color[HTML]{FF0000} \textbf{68.49}}/{\color[HTML]{0070C0} \textbf{1.4207}}& {\color[HTML]{FF0000} \textbf{0.6332}}/{\color[HTML]{0070C0} \textbf{0.0173}}& {\color[HTML]{0070C0} \textbf{0.0183}} \\
\multirow{-10}{*}{DrealSR} & CCSR-S1& 28.33/\color[HTML]{FF0000}{\textbf{0.3391}}                                          & 0.7813/{\color[HTML]{FF0000} \textbf{0.0130}}                                         & 0.3202/{\color[HTML]{FF0000} \textbf{0.0130}}                                          & 0.2327/{\color[HTML]{FF0000} \textbf{0.0079}}                                          & 157.37                                 & 6.82/{\color[HTML]{FF0000} \textbf{0.5352}}& 0.6629/{\color[HTML]{FF0000} \textbf{0.0259}}& {\color[HTML]{0070C0} \textbf{66.21}}/{\color[HTML]{FF0000} \textbf{1.2693}}& 0.6079/{\color[HTML]{FF0000} \textbf{0.0133}}& {\color[HTML]{FF0000} \textbf{0.0140}} \\ \hline
\end{tabular}
}
\label{comparison-1}
\end{table*}

\begin{table*}
\centering
 
 \caption{
 \begin{justify}
        \justifying
        Quantitative comparison among the standard DM-based SR methods with 3 diffusion steps. The best and the second-best results are highlighted in {\color[HTML]{FF0000} \textbf{red}} and  {\color[HTML]{2E75B5} \textbf{blue}}, respectively.
        \end{justify}}

   \resizebox{\linewidth}{!}{
\begin{tabular}{c|c|cccccccccc}
\hline
Datasets                  & Methods     & PSNR/G-STD& SSIM/G-STD& LPIPS/G-STD& DISTS/G-STD& FID                                    & NIQE/G-STD& CLIPIQA/G-STD& MUSIQ/G-STD                                  & MANIQA/G-STD                                  & L-STD                                  \\ \hline
                          & StableSR-S3 & {\color[HTML]{0070C0} \textbf{25.00/0.1894}}          & {\color[HTML]{0070C0} \textbf{0.6304/0.0094}}         & 0.4136/0.0152                                          & 0.3056/0.0104                                   & 42.88                                  & 9.82/0.6786                                           & 0.4706/0.0424                                         & 47.93/{\color[HTML]{0070C0} \textbf{2.5984}}                                 & 0.4405/0.0174                                 & {\color[HTML]{0070C0} \textbf{0.0179}} \\
                          & DiffBIR-S3  & {\color[HTML]{FF0000} \textbf{25.28/0.1825}}          & {\color[HTML]{FF0000} \textbf{0.6346}}/0.0109        & 0.4277/0.0177                                          & 0.3261/0.0123                                          & 54.58                                  & 12.52/1.2433                                          & 0.4014/0.0448                                            & 42.99/3.0343                                 & 0.3707/0.0205                                 & {\color[HTML]{FF0000} \textbf{0.0137}} \\
                          & PASD-S3     & 24.74/0.3186                                          & 0.6287/0.0111                                         & {\color[HTML]{0070C0} \textbf{0.3711}}/0.0220          & 0.2397/0.0120          & 38.74  & 5.78/0.5174         & 0.5855/0.0556                                            & 61.48/3.3128                                 & 0.5444/0.0287                                 & 0.0245                                 \\
                          & SeeSR-S3    & 24.14/0.3791                                          & 0.5877/0.0200                                         & 0.4105/0.0316                                          & 0.2650/0.0137                                          & 45.66                                  & 8.54/0.9152                                           & 0.6864/0.0565            & 63.83/3.4693 & 0.5533/0.0350 & 0.0290  \\
                          
    & CCSR-S2& 24.17/0.2162        &  0.6130/0.0106         & 0.3152/{\color[HTML]{0070C0} \textbf{0.0138}}                                          &{\color[HTML]{0070C0} \textbf{ 0.2216/0.0102}}                                          & {\color[HTML]{0070C0} \textbf{36.08}}                                  &{\color[HTML]{0070C0} \textbf{ 5.62/}}{\color[HTML]{0070C0} \textbf{0.3798}}                                           & {\color[HTML]{FF0000} \textbf{0.7000}}/{\color[HTML]{0070C0} \textbf{0.0378}}            & {\color[HTML]{FF0000} \textbf{71.65}}/{\color[HTML]{FF0000} \textbf{1.1809}} & {\color[HTML]{FF0000} \textbf{0.6480}}/{\color[HTML]{0070C0} \textbf{0.0154}} & 0.0265  \\
    
\multirow{-6}{*}{DIV2K}   & CCSR-S1     & 24.31/0.1932                                          & 0.6283/{\color[HTML]{FF0000} \textbf{0.0082}}                                         & {\color[HTML]{FF0000} \textbf{0.2979/0.0111}}          & {\color[HTML]{FF0000} \textbf{0.2020/0.0083}}          & {\color[HTML]{FF0000} \textbf{30.83}}  & {\color[HTML]{FF0000} \textbf{5.32/0.2982}}           & {\color[HTML]{0070C0} \textbf{0.6754}}/{\color[HTML]{FF0000} \textbf{0.0298}}            & {\color[HTML]{0070C0} \textbf{69.52}}/{\color[HTML]{0070C0} \textbf{1.1905}} & {\color[HTML]{0070C0} \textbf{0.6187}}/{\color[HTML]{FF0000} \textbf{0.0136}} & 0.0201                                 \\ \hline
                          & StableSR-S3 & 26.01/0.2949                                          & {\color[HTML]{0070C0} \textbf{0.7421/0.0092}}         & 0.3475/0.0138                                          & 0.2737/{\color[HTML]{0070C0} \textbf{0.0088 }}                                         & 144.62                                 & 9.19/0.6663                                           & 0.5578/0.0470                                            & 60.51/1.9635                                & 0.5264/0.0183                                 & 0.0154                                 \\
                          & DiffBIR-S3  & {\color[HTML]{FF0000} \textbf{26.65}}/{\color[HTML]{0070C0} \textbf{0.2515}}          & 0.7376/0.0152                                         & 0.3440/0.0177                                          & 0.2952/0.0108                                          & 150.93                                 & 13.51/1.5631                                          & 0.4957/0.0506                                            & 54.44/2.9555                                 & 0.4431/0.0246                                 & {\color[HTML]{0070C0} \textbf{0.0141}} \\
                          & PASD-S3     & {\color[HTML]{0070C0} \textbf{26.59}}/0.4213          & 0.7527/0.0120                                         & 0.3021/0.0156          & {\color[HTML]{0070C0} \textbf{0.2206}}/0.0089          &136.72 &5.95/0.5059          & 0.5678/0.0541                                            & 64.05/2.5824                                 & 0.5683/0.0253                                 & 0.0180                                 \\
                          & SeeSR-S3    & 25.47/0.4674                                          & 0.6909/0.0221                                         & 0.3782/0.0263                                          & 0.2728/0.0111                                          & 144.61                                 & 8.56/1.0359                                           & {\color[HTML]{FF0000} \textbf{0.6848}}/0.0445            &67.21/2.4527 & 0.5922/0.0302 & 0.0256                                 \\
& CCSR-S2& 25.86/0.3032                                          & 0.7335/0.0115& 0.2941/{\color[HTML]{0070C0} \textbf{0.0135}}& 0.2295/0.0096&  {\color[HTML]{0070C0} \textbf{126.12}}                                 & 6.07/ {\color[HTML]{0070C0} \textbf{0.4838}}                                           & 0.6561/{\color[HTML]{0070C0} \textbf{0.0347}}            & {\color[HTML]{FF0000} \textbf{71.17}}/{\color[HTML]{0070C0} \textbf{1.3055}} & {\color[HTML]{FF0000} \textbf{0.6656}}/{\color[HTML]{0070C0} \textbf{0.0156}} & 0.0194 \\
\multirow{-6}{*}{RealSR}  & CCSR-S1     & 25.97/{\color[HTML]{FF0000} \textbf{0.1976}}                                          & {\color[HTML]{FF0000} \textbf{0.7493/0.0070}}         & {\color[HTML]{FF0000} \textbf{0.2804/0.0077}}          & {\color[HTML]{FF0000} \textbf{0.2121/0.0058}}          & {\color[HTML]{FF0000} \textbf{121.43}} & {\color[HTML]{FF0000} \textbf{5.80/0.3474}}           & {\color[HTML]{0070C0} \textbf{0.6278}}/{\color[HTML]{FF0000} \textbf{0.0256}}            & {\color[HTML]{FF0000} \textbf{69.17/0.9194}} & {\color[HTML]{0070C0} \textbf{0.6405}}/{\color[HTML]{FF0000} \textbf{0.0105}} & {\color[HTML]{FF0000} \textbf{0.0140}} \\ \hline
                          & StableSR-S3 & {\color[HTML]{0070C0} \textbf{29.65}}/{\color[HTML]{FF0000} \textbf{0.3388}}          & {\color[HTML]{FF0000} \textbf{0.8064/0.0108}}         & 0.3620/ {\color[HTML]{0070C0} \textbf{0.0174}}                                          & 0.2858/{\color[HTML]{0070C0} \textbf{0.0099  }}                                        & 168.74 & 10.85/0.8078                                          & 0.4565/0.0415                                            & 49.82/2.3432                              & 0.4408/0.0201                                 & {\color[HTML]{0070C0} \textbf{0.0124}} \\
                          & DiffBIR-S3  & {\color[HTML]{FF0000} \textbf{29.67}}/0.3819          & 0.7998/0.0201                                         & 0.3617/0.0264                                          & 0.3189/0.0140                                          & 169.44                                 & 14.41/1.6693                                          & 0.4058/0.0521                                            & 42.77/3.3852                                 & 0.3639/0.0199                                & {\color[HTML]{FF0000} \textbf{0.0120}} \\
                          & PASD-S3     & 29.29/0.4944                                          & {\color[HTML]{0070C0} \textbf{0.8025}}/0.0131         & {\color[HTML]{0070C0} \textbf{0.3279}}/0.0198          & {\color[HTML]{0070C0} \textbf{0.2397}}/0.0113          & 166.03                                 & 7.41/{\color[HTML]{0070C0} \textbf{0.6480}}           & 0.6034/0.0570                                            & 58.19/3.5485                                 & 0.5144/0.0325                                 & 0.0163                                 \\
                          & SeeSR-S3    & 28.42/0.6163                                          & 0.7444/0.0270                                         & 0.4110/0.0350                                          & 0.2981/0.0140                                          & 175.04                                 & 9.83/1.2931                                           & {\color[HTML]{0070C0} \textbf{0.6620}}/0.0580            & 62.11/3.6558 & 0.5359/0.0384 & 0.0239                                 \\
                 & CCSR-S2
& 28.44/0.4365                                     & 0.7724/0.0172                                         & 0.3397/ 0.0181                                        & 0.2563/0.0123                                        & 
       {\color[HTML]{0070C0} \textbf{161.94 }}                                &{\color[HTML]{0070C0} \textbf{ 7.01}}/0.6728& {\color[HTML]{FF0000} \textbf{0.6695}}/{\color[HTML]{0070C0} \textbf{0.0343}}            & {\color[HTML]{FF0000} \textbf{68.49}}/{\color[HTML]{0070C0} \textbf{1.4207}}& {\color[HTML]{FF0000} \textbf{0.6332}}/{\color[HTML]{0070C0} \textbf{0.0173}}& 0.0183 \\
\multirow{-6}{*}{DrealSR} & CCSR-S1     & 28.33/ {\color[HTML]{0070C0} \textbf{0.3391}}                                          & 0.7813/{\color[HTML]{0070C0} \textbf{0.0130  }}                                       & {\color[HTML]{FF0000} \textbf{0.3202/0.0130}}          & {\color[HTML]{FF0000} \textbf{0.2327/0.0079}}          & {\color[HTML]{FF0000} \textbf{157.37}} & {\color[HTML]{FF0000} \textbf{6.82/0.5352}}           & {\color[HTML]{0070C0} \textbf{0.6629}}/{\color[HTML]{FF0000} \textbf{0.0259}}            & {\color[HTML]{0070C0} \textbf{66.21}}/{\color[HTML]{FF0000} \textbf{1.2693}} & {\color[HTML]{0070C0} \textbf{0.6079}}/{\color[HTML]{FF0000} \textbf{0.0133}} & 0.0140                                 \\ \hline
\end{tabular}
}
\label{table_comparison_fewsteps}
\end{table*}

\begin{figure*}
	\centering 
	\includegraphics[scale=1.5]{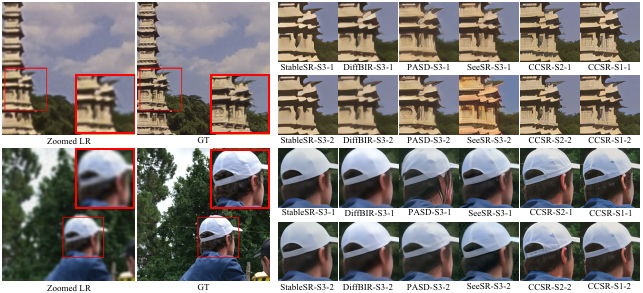}

\caption{Visual comparisons (better zoom-in on screen) between CCSR and state-of-the-art GAN-based and DM-based methods (including StableSR \cite{stablesr}, DiffBIR \cite{diffbir}, PASD \cite{pasd}, SeeSR \cite{seesr} and SUPIR \cite{supir}) with 3 diffusion steps. The SR results become more stable with reduced diffusion steps, but the details become blurry as well.}
\label{compare_fewstep_other}

\end{figure*}

\begin{table*}
\centering

 \caption{
\begin{justify}
        \justifying 
 Quantitative comparison between CCSR and state-of-the-art efficient DM-based SR methods, which require less than 5 diffusion steps, on both synthetic and real-world test datasets. $S$ denotes the number of diffusion steps. Note that the G-STD is not available for FID, because FID measures the statistical distance between two groups of images. The best and the second-best results are highlighted in {\color[HTML]{FF0000} \textbf{red}} and  {\color[HTML]{2E75B5} \textbf{blue}}, respectively.
 \end{justify}}
 
   \resizebox{\linewidth}{!}{
\begin{tabular}{c|c|cccccccccc}
\hline
Datasets & Methods                                                  & PSNR/G-STD& SSIM/G-STD& LPIPS/G-STD& DISTS/G-STD& FID                                    & NIQE/G-STD& CLIPIQA /G-STD& MUSIQ/G-STD                                  & MANIQA/G-STD                                  & L-STD                                  \\ \hline
    & AddSR-S4                                                 & 22.17/0.3712                                          & 0.5273/0.0175                                         & 0.4103/0.0195                                          & 0.2384/0.0100                                          & 35.63                                  & 5.27/0.4106                                           & {\color[HTML]{FF0000} \textbf{0.7499}}/{\color[HTML]{0070C0} \textbf{0.0321}}            & {\color[HTML]{0070C0} \textbf{70.63}}/1.6088 & {\color[HTML]{FF0000} \textbf{0.6604}}/0.0222 & 0.0485                                 \\
         & AddSR-S1                                                 & 23.32/0.3634          & 0.5910/0.0127                                         & 0.3628/0.0174                                          & 0.2124/0.0101                                          & {\color[HTML]{0070C0} \textbf{29.85}}  & {\color[HTML]{0070C0} \textbf{4.76}}/0.4532           & 0.5629/0.0491                                            & 63.31/2.6068                                 & 0.5676/0.0243                                 & 0.0390                                 \\
         & SinSR-S1                                                 & 24.43/0.2706                                          & 0.6012/0.0136                                         & 0.3262/0.0180                                          & 0.2066/{\color[HTML]{0070C0} \textbf{0.0096 }}                                         & 35.45                                  & 6.02/0.4090                                           & 0.6499/0.0458                                            & 62.80/2.0596                                 & 0.5395/0.0152                                 & 0.0368                                 \\
         & OSEDiff-S1                                               & 23.72/-                                                 & 0.6108/-                                                & {\color[HTML]{FF0000} \textbf{0.2941}}/-                 & {\color[HTML]{FF0000} \textbf{0.1976}}/-                 & {\color[HTML]{FF0000} \textbf{26.32}}  & {\color[HTML]{FF0000} \textbf{4.71}}/-                  & 0.6683/-                                                   & 67.97/-                                        & 0.6148/-                                        & -                                      \\
         & CCSR-S2& {\color[HTML]{0070C0} \textbf{24.17/0.2162}}          & {\color[HTML]{0070C0} \textbf{0.6130/0.0106}}         & 0.3152/{\color[HTML]{0070C0} \textbf{0.0138}}                                          & 0.2216/0.0102                                          & 36.08                                  & 5.62/{\color[HTML]{0070C0} \textbf{0.3798}}                                           & {\color[HTML]{0070C0} \textbf{0.7000}}/0.0378            & {\color[HTML]{FF0000} \textbf{71.65}}/{\color[HTML]{FF0000} \textbf{1.1809}} & {\color[HTML]{0070C0} \textbf{0.6480/0.0154}} & {\color[HTML]{0070C0} \textbf{0.0265}} \\
         \multirow{-6}{*}{DIV2K} & CCSR -S1& {\color[HTML]{FF0000} \textbf{24.32/0.1932}}          & 0.6283/{\color[HTML]{FF0000} \textbf{0.0082}}         & {\color[HTML]{0070C0} \textbf{0.2979}} /{\color[HTML]{FF0000} \textbf{0.0111}}         & {\color[HTML]{0070C0} \textbf{0.2020}}/{\color[HTML]{FF0000} \textbf{0.0083}}          & 30.83                                  & 5.32/{\color[HTML]{FF0000} \textbf{0.2982}}                                           & 0.6754/{\color[HTML]{FF0000} \textbf{0.0298}}                                            & 69.52/{\color[HTML]{0070C0} \textbf{1.1905}}                                 & 0.6187/{\color[HTML]{FF0000} \textbf{0.0136}}                                 & {\color[HTML]{FF0000} \textbf{0.0201}} \\ \hline
   & AddSR-S4                                                 & 23.32/0.4117                                          & 0.6397/0.0191                                         & 0.3949/0.0174                                          & 0.2620/0.0092                                          & 151.94                                 & 5.71/0.5409                                           & {\color[HTML]{FF0000} \textbf{0.7164}}/{\color[HTML]{0070C0} \textbf{0.0282}}            & {\color[HTML]{0070C0} \textbf{71.12}}/1.4076 & {\color[HTML]{FF0000} \textbf{0.6817}}/0.0178 & 0.0363                                 \\
         & AddSR-S1                                                 & 24.84/0.3604                                          & 0.7075/0.0117                                         & 0.3100/0.0141                                          & 0.2170/{\color[HTML]{0070C0} \textbf{0.0090}}          & 133.53                                 & {\color[HTML]{FF0000} \textbf{5.53}} /0.5295          & 0.5708/0.0454                                            & 66.55/1.7314                                 & 0.6098/0.0182                                 & 0.0292                                 \\
         & SinSR-S1                                                 & {\color[HTML]{FF0000} \textbf{26.30}}/{\color[HTML]{0070C0} \textbf{0.2539}}          & {\color[HTML]{0070C0} \textbf{0.7354}}/0.0123       & 0.3212/0.0202                                          & 0.2346/0.0084                                          & 137.05                                 & 6.31/{\color[HTML]{0070C0} \textbf{0.4043}}                                           & 0.6204/0.0440                                            & 60.41/1.8421                                 & 0.5389/0.0145                                 & 0.0243                                 \\
         & OSEDiff-S1                                               & 25.15/-                                                 & 0.7341/-                                                & {\color[HTML]{0070C0} \textbf{0.2921}}/-                 & {\color[HTML]{0070C0} \textbf{0.2128}}/-                 & {\color[HTML]{0070C0} \textbf{123.50}} & {\color[HTML]{0070C0} \textbf{5.65}}/-                  & {\color[HTML]{0070C0} \textbf{0.6693}}/-                   & 69.09/-                                        & 0.6339/-                                        & -                                      \\
         & CCSR-S2& 25.86/0.3032                                          & 0.7335/{\color[HTML]{0070C0} \textbf{0.0115}}                                         & 0.2941/{\color[HTML]{0070C0} \textbf{0.0135}}                                          & 0.2295/0.0096                                          & 126.12                                 & 6.07/0.4838                                           & 0.6561/0.0347                                            & {\color[HTML]{FF0000} \textbf{71.17}}/ {\color[HTML]{0070C0} \textbf{1.3055}} & {\color[HTML]{0070C0} \textbf{0.6656/0.0156}} & {\color[HTML]{0070C0} \textbf{0.0194}} \\
        \multirow{-6}{*}{RealSR}   & CCSR -S1& {\color[HTML]{0070C0} \textbf{25.97}}/{\color[HTML]{FF0000} \textbf{0.1976}}          & {\color[HTML]{FF0000} \textbf{0.7493/0.0070}}         & {\color[HTML]{FF0000} \textbf{0.2804/0.0077}}          & {\color[HTML]{FF0000} \textbf{0.2121/0.0058}}          & {\color[HTML]{FF0000} \textbf{121.43}} & 5.80/{\color[HTML]{FF0000} \textbf{0.3474}}                                           & 0.6278/{\color[HTML]{FF0000} \textbf{0.0256}}                                            & 69.17/{\color[HTML]{FF0000} \textbf{0.9194}}                                 & 0.6405/{\color[HTML]{FF0000} \textbf{0.0105}}                                 & {\color[HTML]{FF0000} \textbf{0.0140}} \\ \hline
  & AddSR-S4                                                 & 26.73/0.5458                                          & 0.7104/0.0219                                         & 0.4048/0.0204                                          & 0.2717/0.0106                                          & 163.21                                 & 7.52/0.6492                                           & {\color[HTML]{FF0000} \textbf{0.7180/0.0302}}            & {\color[HTML]{0070C0} \textbf{66.30}}/1.8421 & {\color[HTML]{0070C0} \textbf{0.6290}}/0.0240 & 0.0291                                 \\
         & AddSR-S1                                                 & 27.91/0.4627                                          & 0.7725/{\color[HTML]{0070C0} \textbf{0.0138}}         & 0.3203/{\color[HTML]{0070C0} \textbf{0.0146}}                                          & {\color[HTML]{0070C0} \textbf{0.2249/0.0098}}          & {\color[HTML]{0070C0} \textbf{147.72}} & 6.94/0.6193                                           & 0.6005/0.0412                                            & 60.73/2.3063                                 & 0.5474/0.0218                                 & 0.0241                                 \\
         & SinSR-S1                                                 & {\color[HTML]{0070C0} \textbf{28.41/0.3679}}          & 0.7495/0.0194                                         & 0.3741/0.0287                                          & 0.2488/0.0103                                          & 177.05                                 & 7.02/{\color[HTML]{FF0000} \textbf{0.4339}}                                           & 0.6367/0.0408                                            & 55.34/2.2745                                 & 0.4898/0.0172                                 & 0.0240                                 \\
         & OSEDiff-S1                                               & 27.92/-                                                 &{\color[HTML]{FF0000} \textbf{0.7835}}/-                                               & {\color[HTML]{FF0000} \textbf{0.2968}}/-                 & {\color[HTML]{FF0000} \textbf{0.2165}}/-                 & {\color[HTML]{FF0000} \textbf{135.29}} & {\color[HTML]{FF0000} \textbf{6.49}}/-                  & {\color[HTML]{0070C0} \textbf{0.6963}}/-                   & 64.65/-                                        & 0.5899/-                                        & -                                      \\
         & CCSR-S2& {\color[HTML]{FF0000} \textbf{28.44}}/0.4365          & 0.7724/0.0172                                         & 0.3397/0.0181                                          & 0.2563/0.0123                                          & 161.94                                 & 7.01/0.6728                                           & 0.6695/0.0343                                            & {\color[HTML]{FF0000} \textbf{68.49}}/{\color[HTML]{0070C0} \textbf{1.4207}} & {\color[HTML]{FF0000} \textbf{0.6332}}/{\color[HTML]{0070C0} \textbf{0.0173}} & {\color[HTML]{0070C0} \textbf{0.0183}} \\
         \multirow{-6}{*}{DrealSR}  & CCSR -S1& 28.33/{\color[HTML]{FF0000} \textbf{0.3391}}                                          & {\color[HTML]{0070C0} \textbf{0.7813}}/{\color[HTML]{FF0000} \textbf{0.0130}}         & {\color[HTML]{0070C0} \textbf{0.3202}}/{\color[HTML]{FF0000} \textbf{0.0130}}          & 0.2327/{\color[HTML]{FF0000} \textbf{0.0079 }}                                         & 157.37                                 & {\color[HTML]{0070C0} \textbf{6.82/0.5352}}           & 0.6520/{\color[HTML]{0070C0} \textbf{0.0259}}                                            & 66.21/{\color[HTML]{FF0000} \textbf{1.2693}}                                 & 0.6079/{\color[HTML]{FF0000} \textbf{0.0133}}                                 & {\color[HTML]{FF0000} \textbf{0.0140}} \\ \hline
\end{tabular}
}\label{comparison-2}
\end{table*}

\begin{figure*}
	\centering 
	\includegraphics[scale=1.5]{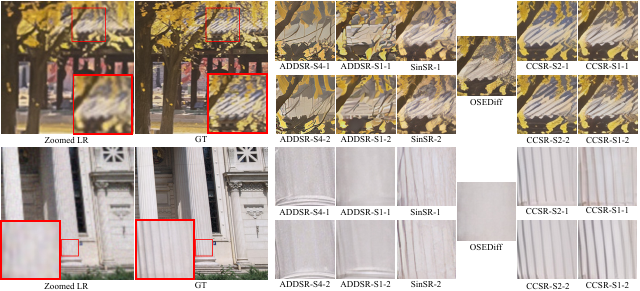}
	\caption{Visual comparisons (better zoom-in on screen) between CCSR and state-of-the-art efficient DM-based SR methods. For each of DM-based methods (except for OSEDiff), two restored images that have the best and worst PSNR values over $10$ runs are shown. Our proposed CCSR works the best to reconstruct accurate structures and realistic, content-consistent and stable details.}
	\label{fig_fewerstep}
\end{figure*}

\begin{table*}\scriptsize
\caption{
\begin{justify}
The inference time and the number of parameters of DM-based SR methods.
\end{justify}
}

\label{inferencetime}
\centering
\resizebox{\linewidth}{!}{
\begin{tabular}{c|cccccc|cccc|cc}
\hline
                     & StableSR& ResShift  & DiffBIR & PASD & SeeSR& SUPIR &AddSR-S4 &AddSR-S1 &SinSR &OSEDiff&CCSR-S2&CCSR-S1 \\ \hline
Inference Steps &    200   &    15      &     50    & 20   &   50 &  50  & 4& 1& 1&1& 2&1\\ \hline
Inference time(s)/Image &   10.03&  0.76& 2.72   &  2.80    & 4.30 & 20.00  & 0.64& 0.21& 0.13&0.12& 0.17&\textbf{0.11}\\ \hline
\#Params(B)&   1.56&     0.18&   1.68&    2.31& 2.51&  18.20  & 2.51& 2.51& 0.18&1.77&1.65&1.65\\ \hline
\end{tabular}}
\end{table*}

\noindent\textbf{Results of Competing Methods with 3 Sampling Steps.} To further show the superiority of CCSR, we run the competing DM-based methods with fewer sampling steps and compare their results on the DIV2K \cite{div2k}, RealSR \cite{realsr} and DrealSR \cite{drealsr} datasets in Table \ref{table_comparison_fewsteps}.  We chose 3 sampling steps for the competing methods because they cannot perform SR reasonably in less than 3 timesteps. Note that SUPIR is not compared since it cannot perform denoising effectively with fewer timesteps, as shown in Fig. \ref{fig1}. We see that on one hand, reducing the sampling steps can curb the uncertainty inherent in the diffusion process, and thus improve the stability and PSNR and SSIM indices of the competing methods (\eg, the L-STD of StableSR on the RealSR dataset improves from 0.0300 to 0.0154). However, all the rest metrics decline for the competing DM-based methods because the reduction of sampling steps reduces their detail generation capability, resulting in deteriorated visual quality of SR outputs. Therefore, simply reducing the sampling steps cannot improves the stability and perceptual quality of existing methods. 

Some visual comparisons are provided in Fig. \ref{compare_fewstep_other}. We have the following observations. 
First, by reducing the sampling steps, the detail generation capability of SatbleSR and DiffBIR is largely reduced, resulting in smoother SR outputs and also suppressing the uncertainty inherent to DMs. Second, there are still noticeable content variations between the two outputs of PASD and SeeSR. Their SR outputs also exhibit visible content differences from the GT, since the fewer diffusion steps cannot support them to generate enough image structures and details.  
In contrast, CCSR can produce more content-consistent structures without sacrificing realistic details.

\subsection{Comparisons with Efficient DM-based SR Methods}

\noindent\textbf{Quantitative Comparisons.} 
We then compare CCSR with those efficient DM-based SR methods, which employ less than five diffusion steps.
The quantitative comparisons across three datasets are presented in Table \ref{comparison-2}.  Despite having fewer diffusion steps, these efficient DM-based SR methods, except for OSEDiff, struggle with instability. Such instability stems from the fact that these methods employ distillation techniques to condense the generation capability of multi-step diffusion models into fewer-step ones, which  inadvertently inherit the instability of their multi-step counterparts. OSEDiff employs a rather different diffusion process. It takes the LR image as the input of the DM without introducing any noise sampling, resulting in a deterministic SR process. However, this approach is hard to be extended to multi-step diffusion, limiting its generation capacity and flexibility for varying perception-distortion requirements. In contrast, CCSR supports both multi-step and one-step DM simultaneously without re-training, accommodating different preferences and requirements. Meanwhile, CCSR shows superior stability performance, as evidenced by its G-STD and L-STD metrics. 

Secondly, the results of AddSR-S4 is biased towards detail generation, resulting in poor performance in reference-based metrics. For example, on the RealSR dataset, the PSNR of AddSR-S4  is 2.65dB lower than that of CCSR-S2. AddSR-S4 shows an advantage in no-reference metrics compared to other efficient methods. However, CCSR-S2 remains competitive with AddSR-S4 on no-reference metrics. When the diffusion steps of AddSR are reduced from 4 to 1, its reference-based metrics improve while the no-reference metrics decline. In contrast, CCSR-S1 exhibits superior performance across both perception and fidelity metrics, striking a good balance between these often conflicting image quality measures.  

Thirdly, SinSR-S1, distilled from ResShift, achieves good full-reference fidelity metrics like PSNR, but its no-reference perception metrics, such as MUSIQ, are poor. This is mainly because ResShift trains a DM from scratch rather than leveraging a pre-trained SD model. Different from SinSR-S1, OSEDiff distills the generative capacity from the pre-trained multi-step SD model, resulting in improved overall performance. When compared to OSEDiff, CCSR-S1 demonstrates superior performance in full-reference fidelity metrics (PSNR/SSIM) while maintaining comparable perception-oriented metrics.  

\noindent\textbf{Qualitative Comparisons.}
Fig.  \ref{fig_fewerstep} provides visual comparisons of the competing efficient DM-based SR methods. As can be seen from the figure, SinSR is difficult to generate details (\eg, the leaves in the first image) due to its under-utilization of pre-trained DM. AddSR-S4 tends to generate unfaithful details. With fewer timesteps, AddSR-S1 produces more faithful results than AddSR-S4 but suffers from blurry details. OSEDiff achieves overall clearer images, but the details of the roof in the first image are compromised. These methods distill the generative capacity from a multi-step pre-trained SD model, yet struggle to control the generative capacity effectively. In contrast, CCSR effectively extracts information from the LR image through a non-uniform sampling strategy and enhances more stable determined details using GAN, enabling the generation of visually pleasing and faithful details.

\subsection{Model Complexity}
\label{sec: complexity}

The number of parameters and the inference time of competing DM-based SR models are listed in Table \ref{inferencetime}. The inference time is calculated on the $\times 4$ SR task with $128 \times 128$ LR images using one NVIDIA A100 80G GPU. 

Among the standard DM-based SR methods, StableSR, DiffBIR and CCSR have similar parameters because they all use the pre-trained SD-2.1-base model with differences in the control part. PASD employs high-level information extractors \cite{YOLO, li2023blip} to extract some high-level information as input to the diffusion network. Therefore, it has more parameters than StableSR, DiffBIR and CCSR. 
SeeSR incorporates the larger RAM (recognize anything model) \cite{RAM} to extract semantic information from LR inputs. This gives it more parameters than PASD. SUPIR employs a more powerful pre-trained model, \eg, SDXL \cite{podell2023sdxl}, striving to achieve higher generation capability. In addition, it adopts the multi-modal LLM LLaVA  \cite{liu2024visual} to extract prompts, resulting in a significantly larger pool of parameters. SUPIR runs the slowest because it is based on SDXL, introduces LLM, and resizes the input LR image to $1024 \times 1024$ for inference. ResShift is trained from scratch and employs substantially smaller parameters with 15 diffusion steps. Therefore, it offers the fastest inference speed but has poor SR quality.

Among the efficient DM-based SR methods, AddSR and SinSR share parameters with their parent models (SeeSR and ResShift, respectively). However, they achieve reduced inference time due to fewer inference steps. SinSR has the fewest parameters, but it struggles to generate fine details. Among those algorithms, OSEDiff stands out with competitive complexity, fewer parameters and shorter inference time. This efficiency is attributed to its use of LoRA for fine-tuning instead of incorporating ControlNet.

CCSR achieves comparable inference time to SinSR although it has larger parameters. This is because the window partition operation is conducted frequently in the Swin Transformer blocks of SinSR, increasing the latency. CCSR does not use additional models to extract high-level information, reducing inference time and parameter count. Therefore, CCSR achieves fewer parameters and faster inference than OSEDiff. Overall, CCSR achieves an excellent balance between model complexity and SR quality.

\section{Conclusion}

To improve the stability of DM-based SR, in this work 
we investigated in-depth how the diffusion priors can help the SR task at different diffusion steps. We found that diffusion priors are more powerful than GANs in generating image main structures when the LR image suffers from significant information loss. However, to further generate high-frequency details, DM may deteriorate the fidelity and go against the goal of image restoration. In contrast, GAN performs favorably well in generating realistic details without changing much the image structures. Based on this observation, we proposed the Content Consistent Super-Resolution (CCSR) approach. Firstly, the coherent structures were generated from the LR image by a diffusion stage. Then, the diffusion process was stopped and the truncated output was sent to the VAE decoder. The VAE decoder was finetuned via adversarial training to acquire the detail enhancement capability without extra computation burden. Extensive experiments demonstrated the superiority of the proposed CCSR method against the existing DM-based methods in SR stability, quality, and efficiency performance.


\ifCLASSOPTIONcaptionsoff
  \newpage
\fi

\footnotesize
\bibliographystyle{IEEEtran}
\bibliography{IEEEabrv,main}

\end{document}